\documentclass[preprint,groupedaddress]{revtex4-1}
\usepackage{graphicx}
\usepackage{dcolumn}
\usepackage{bm}
\usepackage{amsmath}
\usepackage{graphicx}

\newtheorem{theorem}{\hskip\parindent\bf Theorem}
\newtheorem{remark}{\hskip\parindent\bf Remark}

\begin{document}

\preprint{～}

\title[Multiple-parameter bifurcation analysis in a Kuramoto model with time delay and distributed shear]{Multiple-parameter bifurcation analysis in a Kuramoto model with time delay and distributed shear}

\author{Ben Niu}\email{niubenhit@163.com}
\affiliation{Department of  Mathematics, Harbin Institute of Technology (Weihai),\\  Weihai 264209, P.R. China}
\author{Jiaming Zhang, Junjie Wei}
\affiliation{Department of  Mathematics, Harbin Institute of Technology (Weihai),\\  Weihai 264209, P.R. China}
\date{\today}

\begin{abstract}
In this paper, time delay effect and distributed shear are considered in the Kuramoto model. On the Ott-Antonsen's manifold, through analyzing the associated characteristic equation of the reduced functional differential equation, the stability boundary of the incoherent state is derived in multiple-parameter space. Moreover, very rich dynamical behavior such as stability switches inducing synchronization switches can occur in this equation. With the loss of stability, Hopf bifurcating coherent states arise, and the criticality of Hopf bifurcations is determined by applying the normal form theory and the center manifold theorem. On one hand, theoretical analysis indicates that the width of shear distribution and time delay can both eliminate the synchronization then lead the Kuramoto model to incoherence. On the other, time delay can induce several coexisting coherent states.  Finally, some numerical simulations are given to support the obtained results where several bifurcation diagrams are drawn, and the effect of time delay and shear is discussed.
\end{abstract}

\keywords{Kuramoto model;  delay;  bifurcation;  shear;
normal form}
\maketitle

\section{Introduction}
The Kuramoto model was first proposed in \cite{Y. Kuramoto1,Y. Kuramoto2}, consisting of a group of phase oscillators. It is now a mathematical model used most to describe synchronized phenomena \cite{CJP1}. More specifically, it models the behavior of the phase of a large set of weakly coupled, near identical oscillators \cite{S.H. Strogatz1,J.A. Acebron,pla11,pla12}. Its formulation was motivated by the behavior of systems of chemical and biological oscillators, and it has been found widespread applications such as in neuroscience \cite{D. Cumin,M. Breakspear,J. Cabral}, oscillating flame dynamics \cite{G.I. Sivashinsky,D.M. Forrester}, and some physical systems, such as Josphson junctions \cite{S.H. Strogatz2}.

The nonisochronicity, or shear we say, is an important factor in forming spatial-temporal patterns in oscillatory system, which quantifies
the dependence of the  frequency on the amplitude of oscillations, and can also induce very
rich dynamical  behaviors in ensembles of identical or near identical
oscillators \cite{V.Hakim,H.Daido}.
Montbrio and Pazo \cite{E. Montbrio1} have analyzed the Kuramoto model with distributed shear and showed that the onset of collective synchronization is impossible if the width of the shear distribution exceeds a precise threshold. Kuramoto model with shear was also studied in \cite{E. Montbrio2,E. Montbrio3} and the references there in.

Time delay is another important factor in coupled systems. In the real world, every items always takes some time lag to interact with others. As we all know, time delay is ubiquitous in many fields, such as the control theory, ecology, mechanics, management science, physics, neural, etc \cite{J. Hale,K. Goppalsamy,J. Wu}. Recently, there also has been extensive interest in the delay effect on the dynamical behavior of the Kuramoto model \cite{Y. Guo,B. Niu,V. Vlasov,Yeung,M. Choi,J. Cabral,A. Nordenfelt,D. Iatsenko}. So far, Kuramoto model with shear and time delay effect has not been well studied yet.

Motivated by such considerations, we combine time delay effect and shear in a Kuramoto system with the following form
\begin{equation}
\label{1.2}
\dot{\theta}_{j}(t)=\omega_{j}+Kq_{j}+\frac{K}{N}\sum_{k=1}^{N}[\sin(\theta_{k}(t-\tau)-\theta_{j}(t))-q_{j}\cos(\theta_{k}(t-\tau)-\theta_{j}(t))]
\end{equation}
where $\omega_i$'s are randomly distributed frequencies, $K$ is the coupling strength, and $q_j$'s are randomly distributed shears which characterize the impact of  the distance that oscillators move away from the unit circle on the frequencies of the oscillators themselves.

As stated in \cite{E. Montbrio1}, the synchronization transition in \eqref{1.2} fails in case of large spread of shear distribution in the absence of time delay. Also as known to all, time delay will also induce synchronization transition in Kuramoto model\cite{Yeung,M. Choi}. Thus  in this paper we will study the effect of delay together with shear on the synchronization transition from the bifurcation approach. The idea is followed by \cite{benniu} based on the Ott-Antonsen's manifold reduction \cite{E. Ott1,E. Ott2}. Kuramoto model \eqref{1.2} can be reduced into a functional differential equation on a submanifold of the phase space. Trivial equilibrium stands for incoherent state whereas nontrivial periodic oscillations indicate synchronized oscillations of the Kuramoto model. The transition can be described  by Hopf bifurcations which will be discussed in the parameter space consisting of delay and shear.

Through some rigorous  bifurcation analysis, we find that time delay can have very important impact on the system dynamics: compared with the previous results in \cite{E. Montbrio1,E. Montbrio2,E. Montbrio3}, time delay will further prevent synchronization in some cases and will also induce several coexisting coherent states. The bifurcation behavior in multiple parameters spaces are also discussed. These bifurcation phenomena are investigated  both theoretically and numerically in this paper.

The rest of this paper will be  organized as follows. In Section 2, we shall use the Ott-Antonsen method to reduce the original Kuramoto model and obtain the corresponding functional differential equation. In Section 3, we shall consider the stability of the trivial equilibrium standing for the incoherence of Kuramoto model and the existence of the local Hopf bifurcation. In Section 4, the stability and direction of periodic solutions bifurcating from Hopf bifurcations are investigated by using the normal form theory and the center manifold theorem due to \cite{B.D. Hassard}, which gives clearly the location where coherent states appear and their stability. In Section 5, inspired by the method given in \cite{benniu,B. Niu,Corte}, numerical simulations are carried out to support the obtained results.
\section{Ott-Antonsen's manifold reduction}

In this section,  Ott-Antonsen's method is employed to study the delayed Kuramoto model with shear. The approach is actually a direct extension of the method in \cite{E. Ott1,E. Ott2,E. Montbrio1,E. Montbrio2,E. Montbrio3} thus we only give some key steps. To analyze model \eqref{1.2} we adopt its thermodynamic limit $N\rightarrow\infty$, then drop the indices and introduce the probability density for the macroscopical phases $f(\theta,\omega,q,t)$ \cite{S.H. Strogatz3}, which represents the ratio of oscillators with phases between $\theta$ and $\theta+{\rm d}\theta$, natural frequencies between $\omega$ and $\omega+{\rm d}\omega$, and shear between $q$ and $q+{\rm d}q$. Hence the density $f$ must obey the continuity equation
\begin{equation}
\label{2.1}
{\partial}_{t}f=-{\partial}_{\theta}\{\{\omega+Kq+\frac{K}{2{\rm i}}[r(t-\tau){\rm e}^{-{\rm i}\theta(t-\tau)}(1-{\rm i}q)-c.c.]\}f\}
\end{equation}
where c.c. stands for complex conjugate of the preceding term, and the complex-valued order parameter is
\begin{equation}
\label{2.2}
r(t)=\int_{-\infty}^{\infty}\int_{-\infty}^{\infty}\int_{0}^{2\pi}{\rm e}^{{\rm i}\theta}f(\theta,\omega,q,t){\rm d}\theta {\rm d}\omega {\rm d}q
\end{equation}
If the phases are uniformly distributed, we know $r(t)$ vanishes to zero. This state is customarily referred to as incoherence or incoherent state. Similarly $|r(t)|=1$ stands for the completely synchronized state, and $|r(t)|\in(0,1)$  partially synchronized states or coherent state. Since $f(\theta,\omega,q,t)$ is a real function and $2\pi$-periodic in the $\theta$ variable, we know it admits the Fourier expansion
\begin{equation}
\label{2.3}
f(\theta,\omega,q,t)=\frac{p(\omega,q)}{2{\rm \pi}}\sum\limits_{l=-\infty}^{\infty}{f}_{l}(\omega,q,t){\rm e}^{{\rm i}l\theta}
\end{equation}
where $f_l=f_{-l}^*$, $f_0=1$, the $"*"$ stands for the complex conjugate, and $p(\omega,q)$ is the joint probability density function of $\omega$ and $q$. The first order Fourier term $f_1$ is important because it determines the order parameter
\begin{equation}
\label{2.4}
{r}^{*}(t)=\int_{-\infty}^{\infty}\int_{-\infty}^{\infty}p(\omega,q){f}_{1}(\omega,q,t){\rm d}\omega{\rm d}q.
\end{equation}
Inserting the Fourier series \eqref{2.3} into the continuity equation we obtain the following equation
\begin{equation}
\label{2.5}
{\partial}_{t}{f}_{l}=-{\rm i}l(\omega+Kq){f}_{l}+\frac{Kl}{2}[{r}^{*}(t-\tau)(1+{\rm i}q){f}_{l-1}-r(t-\tau)(1-{\rm i}q){f}_{l+1}]
\end{equation}which governs the  dynamics of Kuramoto model.
Ott and Antonsen \cite{E. Ott1,E. Ott2} found that the ansatz
\begin{equation}
\label{2.6}
{f}_{l}(\omega,q,t)=\alpha(\omega,q,t)^{l}
\end{equation}
is a particular and important solution of the Kuramoto model which can eliminate the $\omega$-variable without loss of generality.

In this paper, we further assume the frequency and the shear are independent and both obey the Lorentzian distribution with mean $\omega_{0}$ and $q_{0}$, and the width $\delta$ and $\gamma$, respectively. That is, $g_{1}(\omega)=\frac{\delta/\pi}{(\omega-\omega_{0})^2+\delta^2}$, $g_{2}(q)=\frac{\gamma/\pi}{(q-q_{0})^2+\gamma^2}$, and $p(\omega,q)=g_{1}(\omega)g_{2}(q)$. Integrating \eqref{2.4} with residue theorem we have  $r^*(t)=\alpha(\omega,q,t)$. Substitute $\omega=\omega_0-{\rm i}\delta$ and $q=q_0-{\rm i}\gamma$ into Eq.\eqref{2.5}, then we can get the reduced equation.
\begin{equation}
\label{2.8}
\begin{aligned}
\dot{r}(t)={\rm i}[\omega_0+{\rm i}\delta+K(q_0+{\rm i}\gamma)]r(t)+\frac{K}{2}\{r(t-\tau)[1-{\rm i}(q_0+{\rm i}\gamma)]-r^*(t-\tau)[1+{\rm i}(q_0+{\rm i}\gamma)]r^2(t)\}
\end{aligned}
\end{equation}
This is a delay differential equation, which characterizes the dynamical behavior of \eqref{1.2}, as proved in \cite{E. Ott1,E. Ott2}. Investigating stability conditions for the incoherent state $r(t)=0$ and exploring the existence of Hopf bifurcations will be used to detect the appearence of coherent states.
\section{Stability and bifurcation analysis}
The characteristic equation of the linearization of \eqref{2.8}, around the trivial equilibrium (incoherence), is,
\begin{equation}
\label{3.2}
\lambda+(\delta+K\gamma)-\frac{K}{2}(1+\gamma){\rm e}^{-\lambda\tau}-{\rm i}({\omega}_{0}+K{q}_{0})+{\rm i}\frac{K{q}_{0}}{2}{\rm e}^{-\lambda\tau}=0
\end{equation}
Based on the stability theory of the functional differential equations, if all the roots of Eq.\eqref{3.2} have negative real parts, then the zero solution of system \eqref{2.8} is always asymptotically stable, which means that the Kuramoto model will remain in a incoherent state in the sense of small perturbations. If system \eqref{2.8} has  an orbitally asymptotically stable periodic solution which is bifurcated from Hopf bifurcation after Eq.\eqref{3.2} has purely imaginary roots, then the Kuramoto model will exhibit partial synchronization (coherent states). This Hopf bifurcation gives the critical value of the synchronization transition of the Kuramoto model.

When $\tau=0$, Eq.\eqref{3.2} becomes
$
\lambda-[\frac K2(1-\gamma)-\delta]-{\rm i}(\omega_0+\frac{Kq_0}{2})=0
$.
Without of loss generality, we always consider the case with $\frac K2(1-\gamma)\neq\delta$  such that  zero is not a root of \eqref{3.2}, then the transition is via Hopf bifurcations.
Obviously,
the root of Eq.\eqref{3.2} with $\tau=0$ has negative real part when $\frac K2(1-\gamma)<\delta$, and positive real part when $\frac K2(1-\gamma)>\delta$.

If $\tau>0$, suppose that ${\rm i}\beta$($\beta\neq0$) is a root of Eq.\eqref{3.2}.  Substituting ${\rm i}\beta$ into Eq.\eqref{3.2} and separating the real and imaginary parts yield
\begin{equation}
\begin{aligned}
\label{3.3}
\frac{K}{2}(1+\gamma)\cos\beta\tau-\frac{Kq_0}{2}\sin\beta\tau=&\delta+K\gamma\\
\frac{Kq_0}{2}\cos\beta\tau+\frac{K}{2}(1+\gamma)\sin\beta\tau=&\omega_0+Kq_0-\beta
\end{aligned}
\end{equation}
Obviously, we have $\beta$ can be solved by
\begin{equation*}
\beta=\beta_{\pm}=\omega_0+Kq_0\pm\sqrt{\frac{K^2}{4}[(1+\gamma)^2+q_0^2]-(\delta+K\gamma)^2}
\end{equation*}
Choose $\tau=\tau_j^\pm>0, j=0,1,\cdots$, such that $|\tau_j^\pm\beta_\pm|\in[2j\pi,2(j+1)\pi)$, and (\ref{3.3}) holds true.
Let $\lambda(\tau)=\alpha(\tau)+{\rm i}\beta(\tau)$ be the root of Eq.\eqref{3.2} satisfying $\alpha(\tau_j^\pm)=0$ and $\beta(\tau_j^\pm)=\beta_\pm$.

To investigate the roots' distribution, we also need to verify the  transversality condition. In fact, we have
\begin{equation}\label{transv}
{\alpha}'({\tau}^{+}_{j})
\begin{cases}
>0,~if~{\beta}_{+}>0\\
<0,~if~{\beta}_{+}<0
\end{cases}
\quad\text{and}\quad
{\alpha}'({\tau}^{-}_{j})
\begin{cases}
<0,~if~{\beta}_{-}>0\\
>0,~if~{\beta}_{-}<0
\end{cases}
\end{equation}
This can be proved by
substituting $\lambda(\tau)$ into Eq.\eqref{3.2} and taking the derivative with respect to $\tau$ at $\tau=\tau_j^\pm$, respectively. In fact,
\begin{equation*}
{\alpha}'({\tau}_{j}^{\pm})=\frac{{\beta}_{\pm}({\beta}_{\pm}-\omega_0-Kq_0)}{[1+\tau_j^\pm (\delta+K\gamma)]^2+[\tau_j^\pm({\beta}_{\pm}-\omega_0-Kq_0)]^{2}}
\end{equation*}
Together with the fact $\beta_+-\omega_0-Kq_0>0$ and $\beta_--\omega_0-Kq_0<0$, the transversality condition is verified.

From the discussion above we know that, under the assumption $\frac{K^2}{4}[(1+\gamma)^2+q_0^2]-(\delta+K\gamma)^2>0$, there are two sequences of critical values of $\tau$, $\{\tau_j^+\}$ and $\{\tau_j^-\}$, where Eq.(3.2) has a purely imaginary root when $\tau=\tau_j^\pm$, respectively. We reorder $\{\tau_j^+\}\bigcup\{\tau_j^-\}$ as $\{\tau_j\}$ so that $\tau_0<\tau_1<\tau_2<\ldots.$ Clearly, $\tau_0={\rm min}\{\tau_0^+,\tau_0^-\}$.

In order to state the stability and Hopf bifurcation results of Eq.(\ref{3.2}) clearly,  we make the following assumptions
 \begin{description}
\item [(P1)] $(\delta+K\gamma)^2-\frac{K^2}{4}[(1+\gamma)^2+q_0^2]>0$
\item [(P2)] $(\delta+K\gamma)^2-\frac{K^2}{4}[(1+\gamma)^2+q_0^2]<0$
\item [(P3)] $(\delta+K\gamma)^2+(\omega_0+Kq_0)^2-\frac{K^2}{4}[(1+\gamma)^2+q_0^2]\leq0$
\item [(P4)] $(\delta+K\gamma)^2+(\omega_0+Kq_0)^2-\frac{K^2}{4}[(1+\gamma)^2+q_0^2]>0$
\end{description}

In fact, when investigating the effect of shear, assumptions (P1) can be cast into a different form, $[(1+\gamma)^2+q_0^2]< 4\left(\frac{\delta}K+\gamma\right)^2$. Thus, for fixed   $\gamma$, (P1) means that we are using a small $q_0$, the mean value of shear, or a small $K$, the coupling strength,  or a large $\delta$, the spread of the frequency distribution, whereas (P2) means the contrary. Furthermore, if (P1) holds, so does (P4). Similarly if (P3) holds true,  so does (P2). Thus three cases are considered in the following part: (P1) holds, (P3) holds, and (P2,P4) holds, respectively. In the coming part, we will show that these three cases correspond to that the Kuramoto model exhibits incoherent states, that the Kuramoto model is synchronized when delay exceeds a threshold, and that the model exhibits synchronization windows, i.e., incoherent and coherent states  are interlacing.

Summarizing the above analysis, from the result of  \cite{S. Ruan,J. Wei}, we have
\begin{theorem}
\label{lem3.3}

 \begin{enumerate}
\item Suppose $\frac K2(1-\gamma)<\delta$ holds true,\label{lem3.3-1}
 \begin{enumerate}
 \item \label{lem3.3-1-a}If (P1) holds true, then all the roots of Eq.\eqref{3.2} have negative real parts, and  the zero solution of system \eqref{2.8} is always asymptotically stable, for $\tau>0$.
\item \label{lem3.3-1-b} If  (P3) holds true, then there exists $\tau_0>0$ such that all the roots of Eq.\eqref{3.2} have negative real parts  and  the zero solution of system \eqref{2.8} is  asymptotically stable when $\tau\in[0,\tau_0)$; Eq.\eqref{3.2} has at least one root with positive real part and  the zero solution of system \eqref{2.8} is  unstable  when $\tau>\tau_0$.
\item \label{lem3.3-1-c} If (P2) and (P4) hold, then exists  an integer $n$ such that all the roots of Eq.\eqref{3.2} have negative real parts and  the zero solution of system \eqref{2.8} is  asymptotically stable  when $\tau\in[0,\tau_0)\bigcup\left(\bigcup\limits_{j=1}^{n}(\tau_{2j-1},\tau_{2j})\right)$;  Eq.\eqref{3.2} has at least one root with
positive real part  and  the zero solution of system \eqref{2.8} is  unstable when $\tau\in\left(\bigcup\limits_{j=0}^{n-1}(\tau_{2j},\tau_{2j+1})\right)\bigcup(\tau_{2n},\infty)$.
\end{enumerate}\label{lem3.3-2}
\item Suppose that $\frac K2(1-\gamma)>\delta$ is satisfied, if (P2) and (P4) hold, Eq.\eqref{3.2} has a root with positive real part  and  the zero solution of system \eqref{2.8} is  unstable when $\tau\in[0,\tau_0) $. More precisely, we have
\begin{enumerate}
\item \label{lem3.3-2-a}If $\tau_0=\tau_0^+$, then Eq.\eqref{3.2} has at least one root with positive real part  and  the zero solution of system \eqref{2.8} is  unstable for all $\tau\geq0$.
\item \label{lem3.3-2-b}If $\tau_0=\tau_0^-$, then there exists an integer $n$ such that all the roots of Eq.\eqref{3.2} have negative real parts and  the zero solution of system \eqref{2.8} is  asymptotically stable  when $\tau\in\bigcup\limits_{j=0}^{n-1}(\tau_{2j},\tau_{2j+1})$; Eq.\eqref{3.2} has at least one root with positive real part  and  the zero solution of system \eqref{2.8} is  unstable when $\tau\in[0,\tau_0)\bigcup\left(\bigcup\limits_{j=1}^{n-1}(\tau_{2j-1},\tau_{2j})\right)$ $\bigcup(\tau_{2n-1},\infty).$
\end{enumerate}

\end{enumerate}

Moreover, the system \eqref{2.8} undergoes a Hopf bifurcation at the origin when $\tau=\tau_{j},j=0,1,\ldots$.
\end{theorem}

\begin{figure}
\centering
a)
\includegraphics[width=0.28\textwidth]{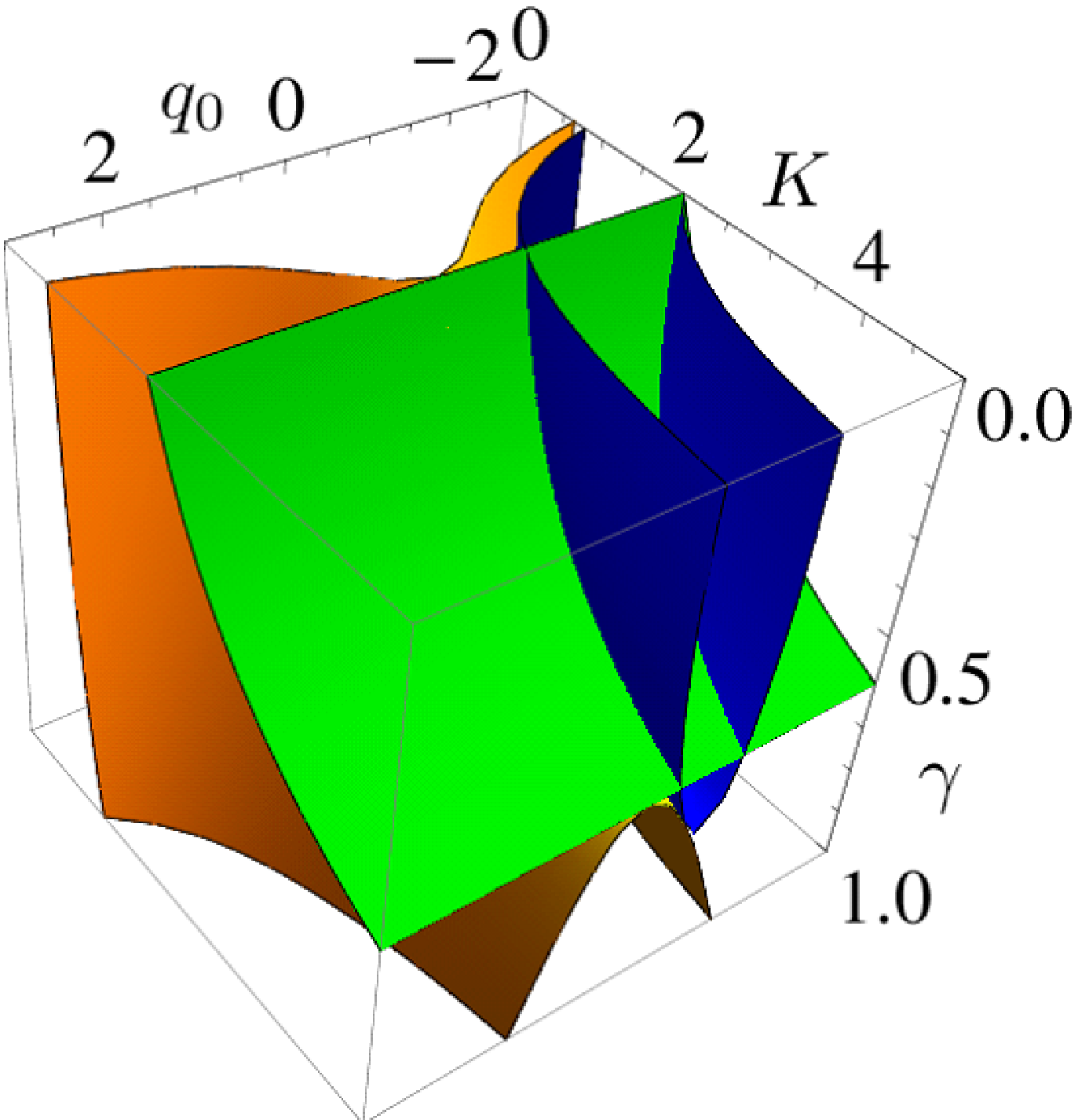}
b)
\includegraphics[width=0.28\textwidth]{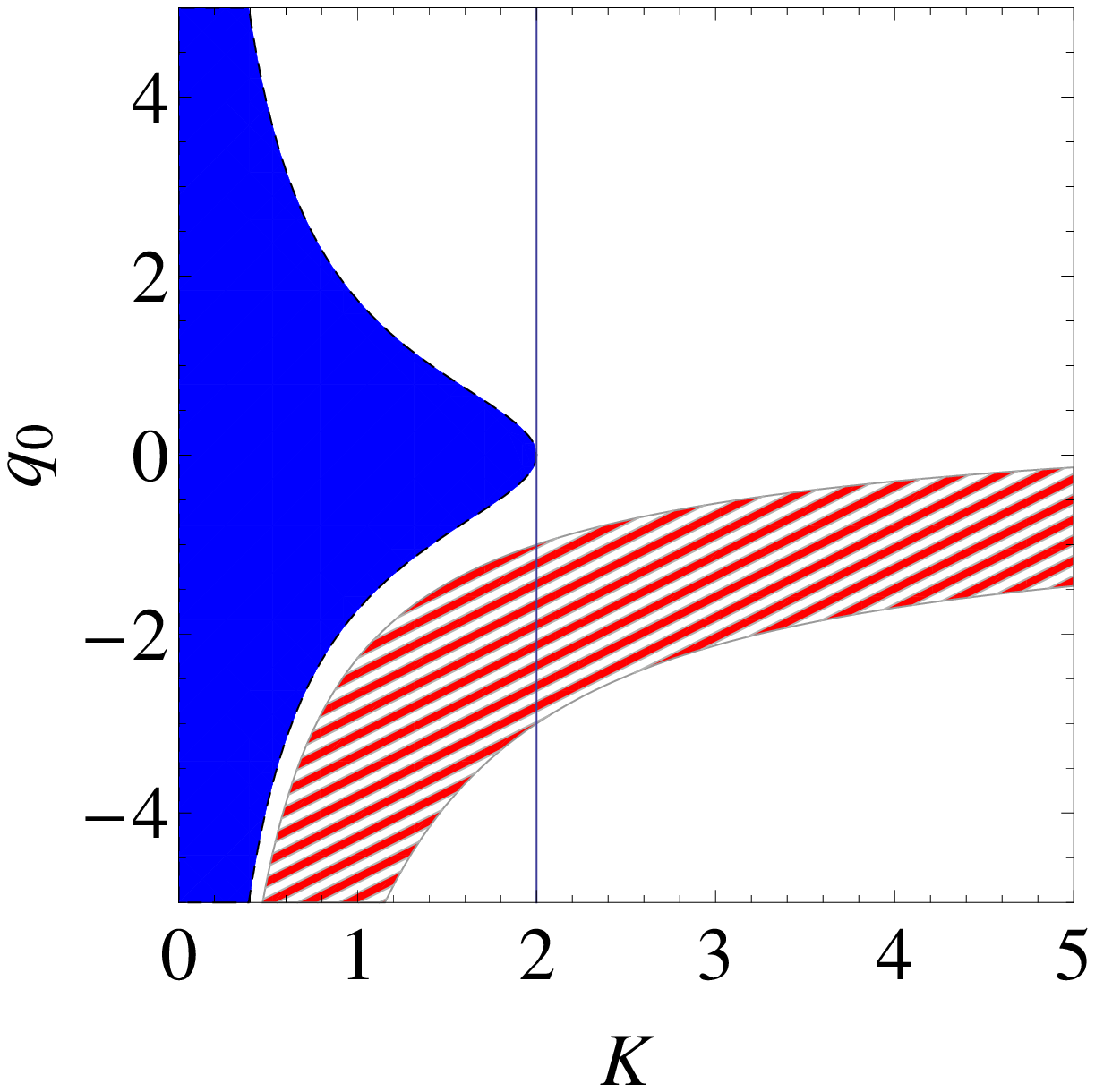}
c)
\includegraphics[width=0.28\textwidth]{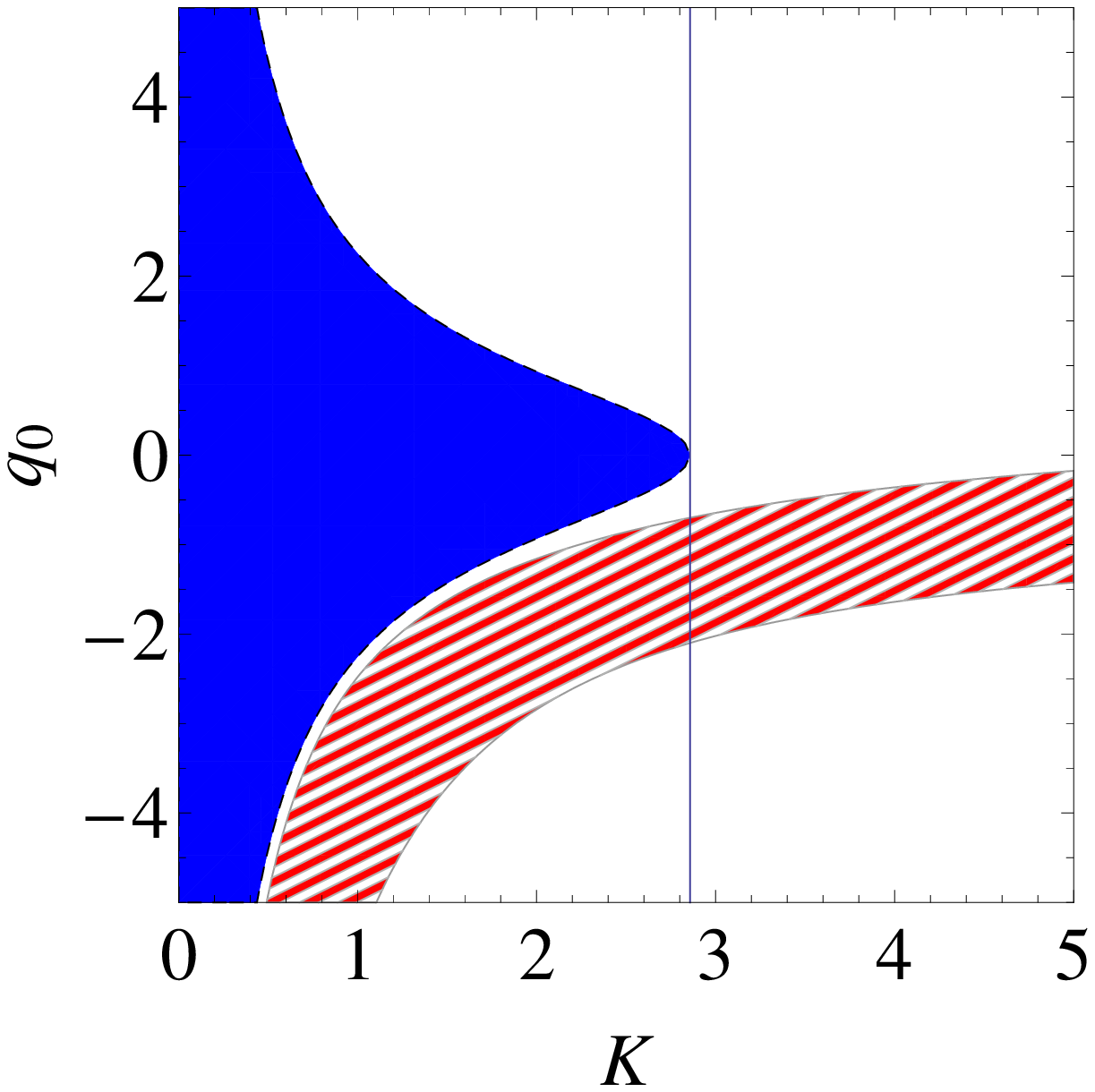}\\
d)
\includegraphics[width=0.28\textwidth]{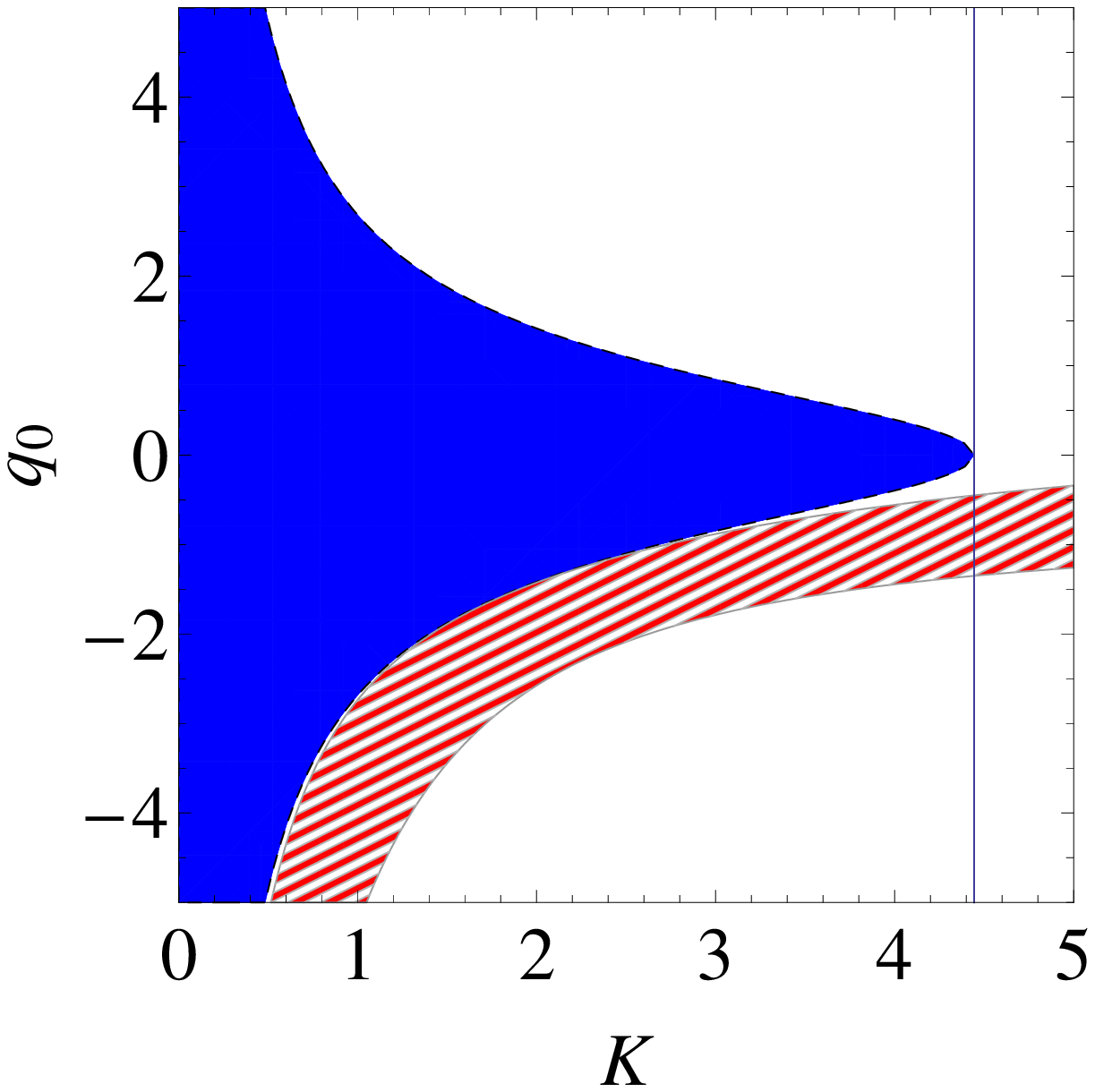}
e)
\includegraphics[width=0.28\textwidth]{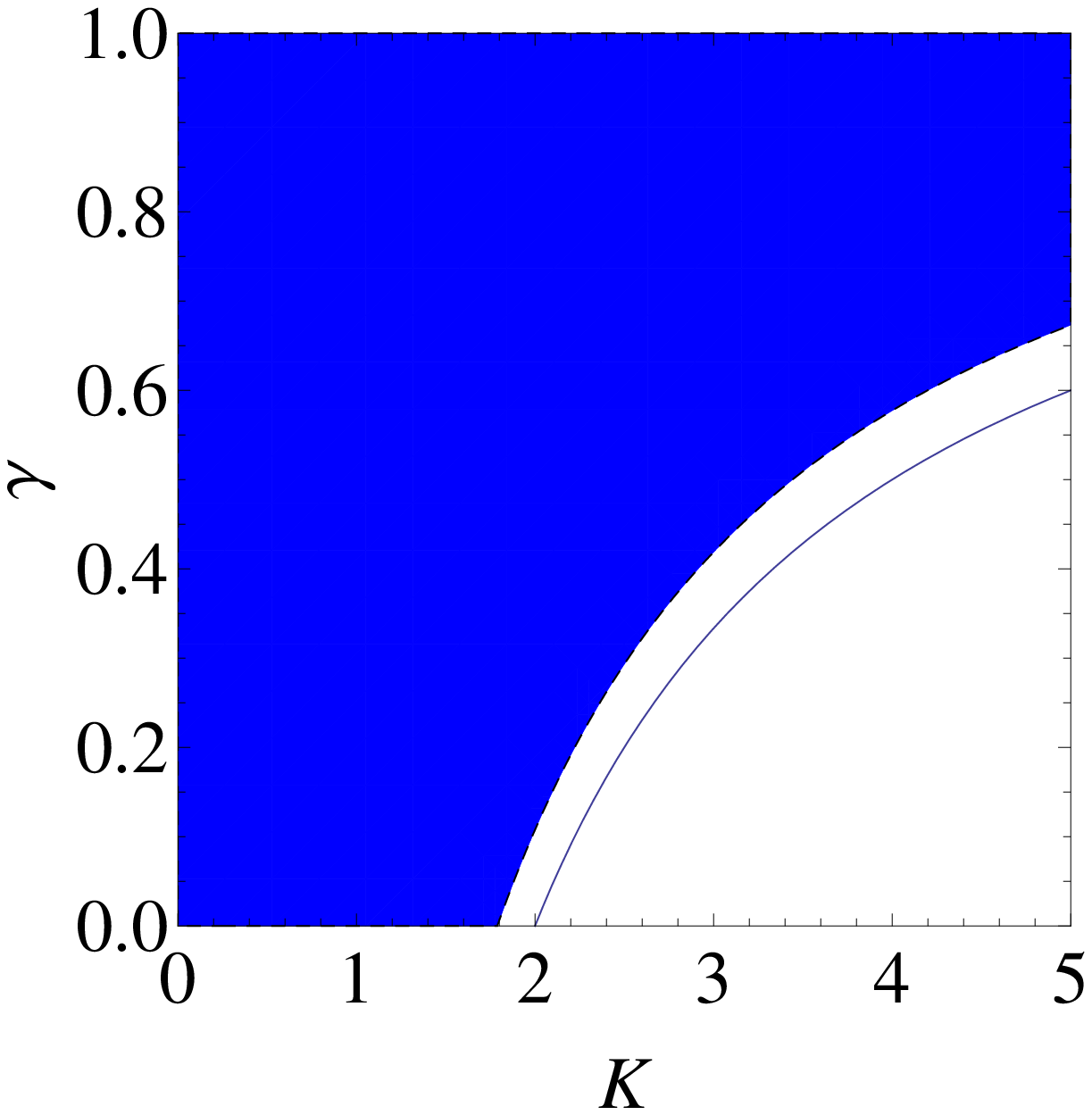}
f)
\includegraphics[width=0.28\textwidth]{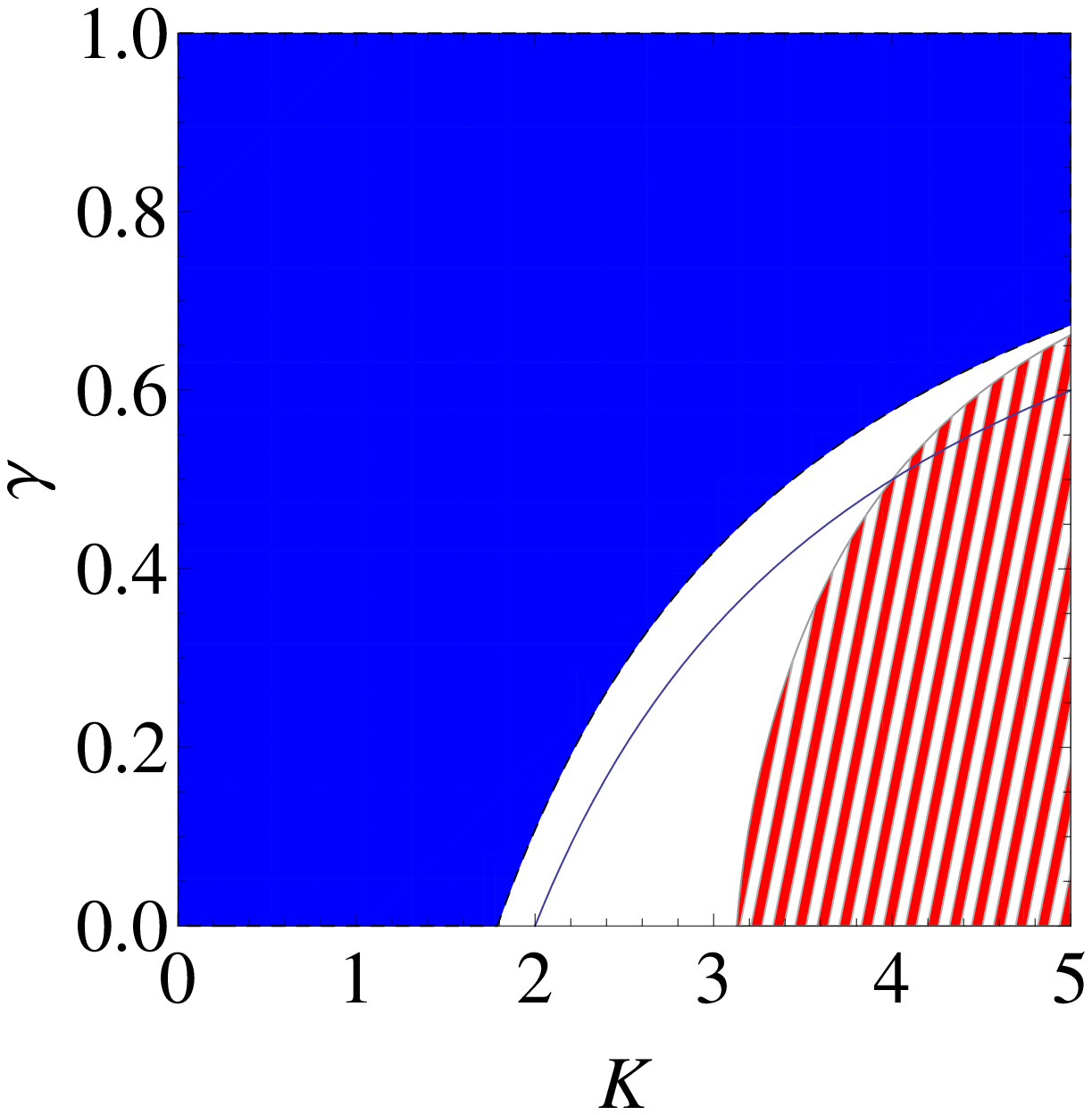}
\caption{For $\omega_0=3$, $\delta=1$,   bifurcation results given in theorem \ref{lem3.3} are illustrated in the original parameters. (a)  critical surfaces in $q_0-K-\gamma$ space. Green surface stands for $\frac K2(1-\gamma)=\delta$, yellow surface $\frac{K^2}{4}[(1+\gamma)^2+q_0^2]-(\delta+K\gamma)^2=0$, and blue surface $(\delta+K\gamma)^2+(\omega_0+Kq_0)^2-\frac{K^2}{4}[(1+\gamma)^2+q_0^2]=0$. (b-f): The critical curve $\frac K2(1-\gamma)=\delta$ (the black curve), the region $\frac{K^2}{4}[(1+\gamma)^2+q_0^2]-(\delta+K\gamma)^2<0$ (color blue) and the region $(\delta+K\gamma)^2+(\omega_0+Kq_0)^2-\frac{K^2}{4}[(1+\gamma)^2+q_0^2]<0$ (color white-red). The parameters in  (b-d) are $\gamma=0,0.3,0.55$, and  in (e-f), $q_0=0.5,-0.5$}\label{newfig}
\end{figure}

\begin{remark}\label{remnew}The conditions (P1)-(P4) given in Theorem \ref{lem3.3} are expressed by all parameters of Eq.(\ref{2.8}). However, these conditions contains square  terms of each parameter, thus, the inequalities (P1)-(P4) cannot be expressed by simple forms. In fact, we can show these conditions graphically in Figure \ref{newfig}.  In Figure \ref{newfig} (b)-(f), on the left side of the curve $\frac K2(1-\gamma)=\delta$, Theorem \ref{lem3.3}-1 applies: the region is divided into three parts. The blue region is the absolutely stable region , i.e., for any time delay the incoherence is stable; in the white-red region the incoherence is stable for small delay and synchronization transition appears at a certain  critical value $\tau_0$; the white part indicates synchronization switches as shown in Theorem \ref{lem3.3}(1)(c). The absolutely stable region becomes larger as increasing the spread $\gamma$. On the right side of the curve $\frac K2(1-\gamma)=\delta$, Theorem \ref{lem3.3}-2 applies: the incoherent state is unstable for $\tau=0$ and  there exists a region (colored white-red) where  synchronization switches may appear as stated in Theorem \ref{lem3.3} (2)(b). The bifurcation conditions are shown as surfaces in $\gamma-K-q_0$ space in Figure \ref{newfig} (a).
\end{remark}

\section{Direction and stability of Hopf bifurcation}

In Section 3, we have obtained the stability  of the incoherent state and a group of conditions which guarantee that the equation undergoes Hopf bifurcation at some critical values of $\tau$. In this section, we shall study the direction and stability of the bifurcating periodic solutions. The method we used is based on the normal form method and the center manifold theory presented in Hassard et al \cite{B.D. Hassard}. We shall compute the center manifold construction of system \eqref{2.8} at $\tau=\bar\tau\in\{\tau_j^\pm,j=0,1,\ldots$\} while the characteristic equation has a root $i\beta_0$.

The following variable can be calculated as shown in the Appendix:
\begin{equation*}
\begin{split}
&c_1(0)=\frac{\rm i}{2\beta_0}[g_{11}g_{20}-2|g_{11}|^2-\frac{|g_{02}|^2}{3}]+\frac{g_{21}}{2}\\
&\mu_2=-\frac{\mathrm{Re}c_1(0)}{\alpha^{'}(\overline{\tau})}\\
&\beta_2=2\mathrm{Re}c_1(0)\\
&T_2=-\frac{\mathrm{Im}c_1(0)+\mu_2\beta^{'}(\overline{\tau})}{\beta_0}
\end{split}
\end{equation*}

 Based on  \cite{B.D. Hassard} and \cite{S. Wiggein}, we have the following results.
\begin{theorem}
\label{the4.1}
If $\mu_2>0$ {\rm (}resp. $<0${\rm )}, then the direction of the Hopf bifurcation of system \eqref{4.1} at the origin when $\tau=\overline{\tau}$ is supercritical {\rm (}resp. subcritical{\rm )}. If $\beta_2<0$ {\rm (}resp. $>0${\rm )}, then the bifurcating  solution on the center
manifold is asymptotically {\rm (}resp. unstable{\rm )}. If $T_2>0$ {\rm (}resp. $<0${\rm )}, the period of the bifurcating periodic solutions increases {\rm (}resp. decreases{\rm )}.
\end{theorem}

\begin{remark}
Assume the incoherence is stable when $\tau=0$, and Hopf bifurcation occurs at $\bar \tau$. By using the global Hopf bifurcation results (see \cite{benniu,JH. Wu}), we know there will always exist hysteresis loop near the subcritical bifurcations. This will be shown in the simulation part such as  Figure \ref{fig2}(a).
\end{remark}

\section{Numerical examples}

In this section, with the aid of  theoretical results obtained in the previous sections, we will carry out several groups of numerical simulations to illustrate the complicated dynamical behavior of the Kuramoto model with time delay and shear. All the illustrations we will show are about the bifurcation branches of the model, because   bifurcation branches are widely used in literatures \cite{benniu,B. Niu,Corte}, which are also a quite unambiguous way to show the change of numbers of steady states or periodic oscillations.
\subsection{Single-parameter bifurcations}

We first give some numerical simulations about the reduced model \eqref{2.8} when varying one parameter.

\begin{figure}\centering
a)
\includegraphics[width=0.46\textwidth]{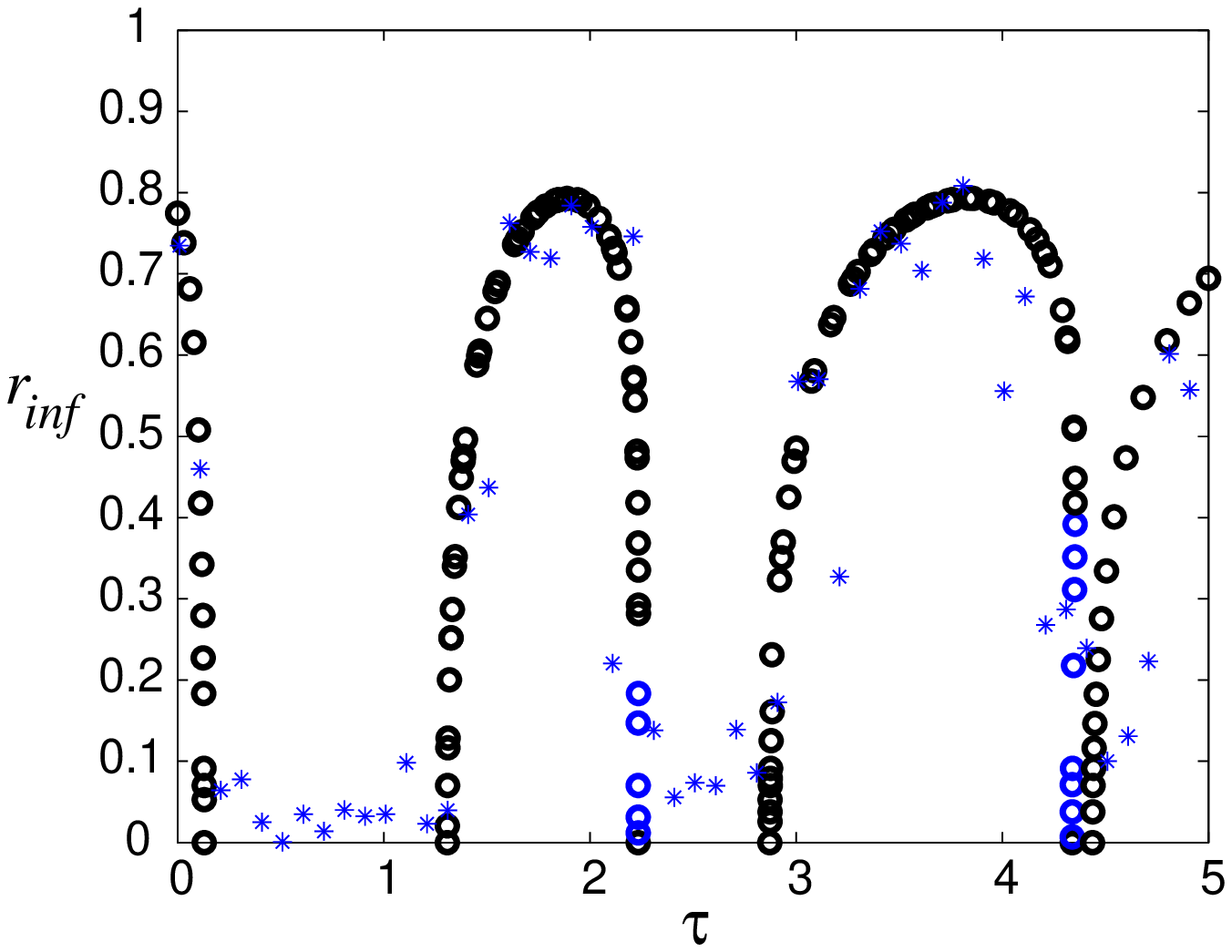}
b)
\includegraphics[width=0.46\textwidth]{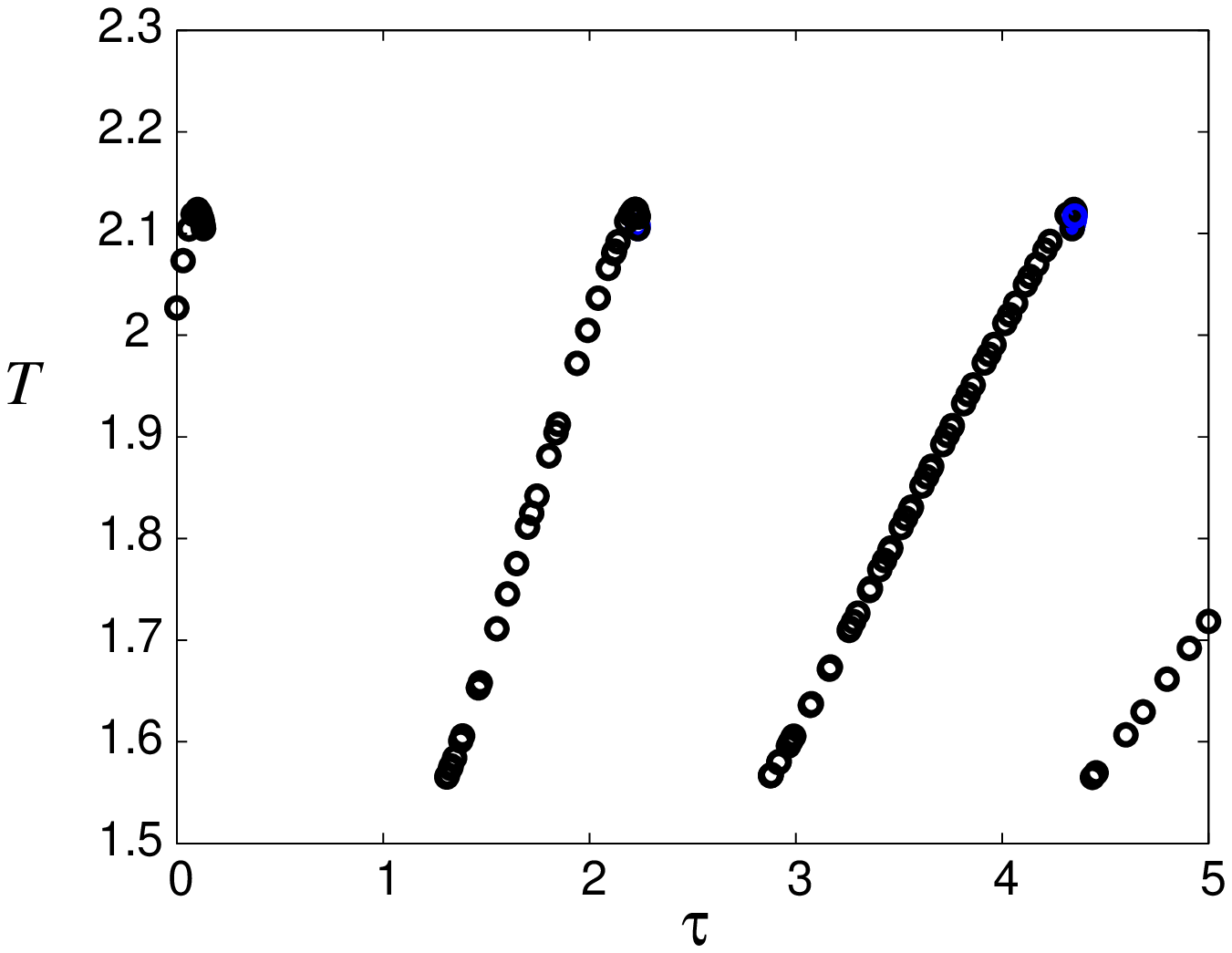}
\caption{(a) The bifurcation diagram of Eq.\eqref{2.8}. (b) The period of the corresponding  bifurcating solutions of Eq.\eqref{2.8}. Here we use $K=1$, $\omega_{0}=3$, $\delta=0.1$, $q_{0}=0.5$, $\gamma=0.5$ and $r_{inf}$ stands for $|r(\infty)|$. Stable periodic orbits are labeled by black circles and unstable (1 Floquet exponent with positive real part) periodic orbits are labeled by blue circles. The order parameters by integrating the Kuramoto model \eqref{1.2} with $N=500$ are marked by blue stars.}\label{fig1}
\end{figure}

Choosing $K=1$, $\omega_{0}=3$, $\delta=0.1$, $q_{0}=0.5$, $\gamma=0.5$, which satisfies the condition of Theorem \ref{lem3.3}(2)(b), and letting $\tau$ vary, we have a bifurcation diagram shown in Figure \ref{fig1}(a). By using DDE-Biftool \cite{K. Engelborghs1, K. Engelborghs2} and computing the numbers of Floquet exponents of periodic solutions with positive real part, the order paramters $r_{inf}=|r(\infty)|$ and stability of periodic solutions of \eqref{2.8} (i.e., stability of the coherent states of  \eqref{1.2}) are shown.  Stable periodic orbits are labeled by black circles, unstable (1/2/3/4 Floquet exponents with positive real part) periodic orbits are labeled by blue/cyan/red/magenta circles,  respectively. The order parameters by integrating the Kuramoto model \eqref{1.2} with $N=500$ are marked by blue stars.  When $\tau=0$, $|r(\infty)|\approx0.8$, the system \eqref{2.8} exhibits partially coherent state. As $\tau$ increases,  $|r(\infty)|$ first reduces to 0, which means the system becomes incoherent, then the system switches between the coherent state and the incoherent state, i.e., the synchronization switches appear. We find it supports results in the Theorem \ref{lem3.3}(2)(b). In Figure \ref{fig1}(b) we calculate the period of the branches of bifurcating periodic solutions, where we find the period of the bifurcating synchronized oscillations increases as time delay increases. Some numerical calculations yield that $T_2>0$, which means that these numerical simulations verify the results in Theorem \ref{the4.1}.

\begin{figure}\centering
a)
\includegraphics[width=0.46\textwidth]{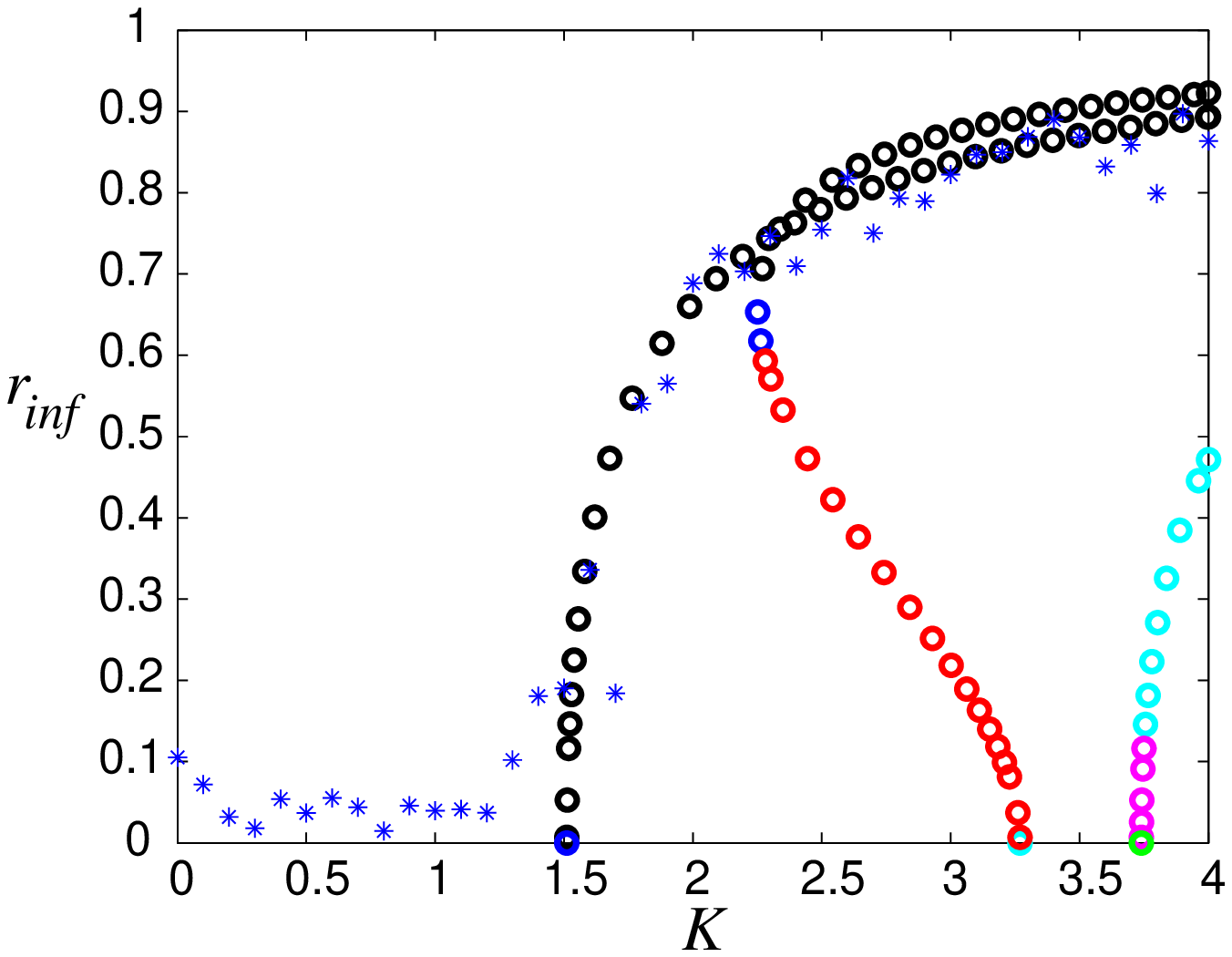}
b)
\includegraphics[width=0.46\textwidth]{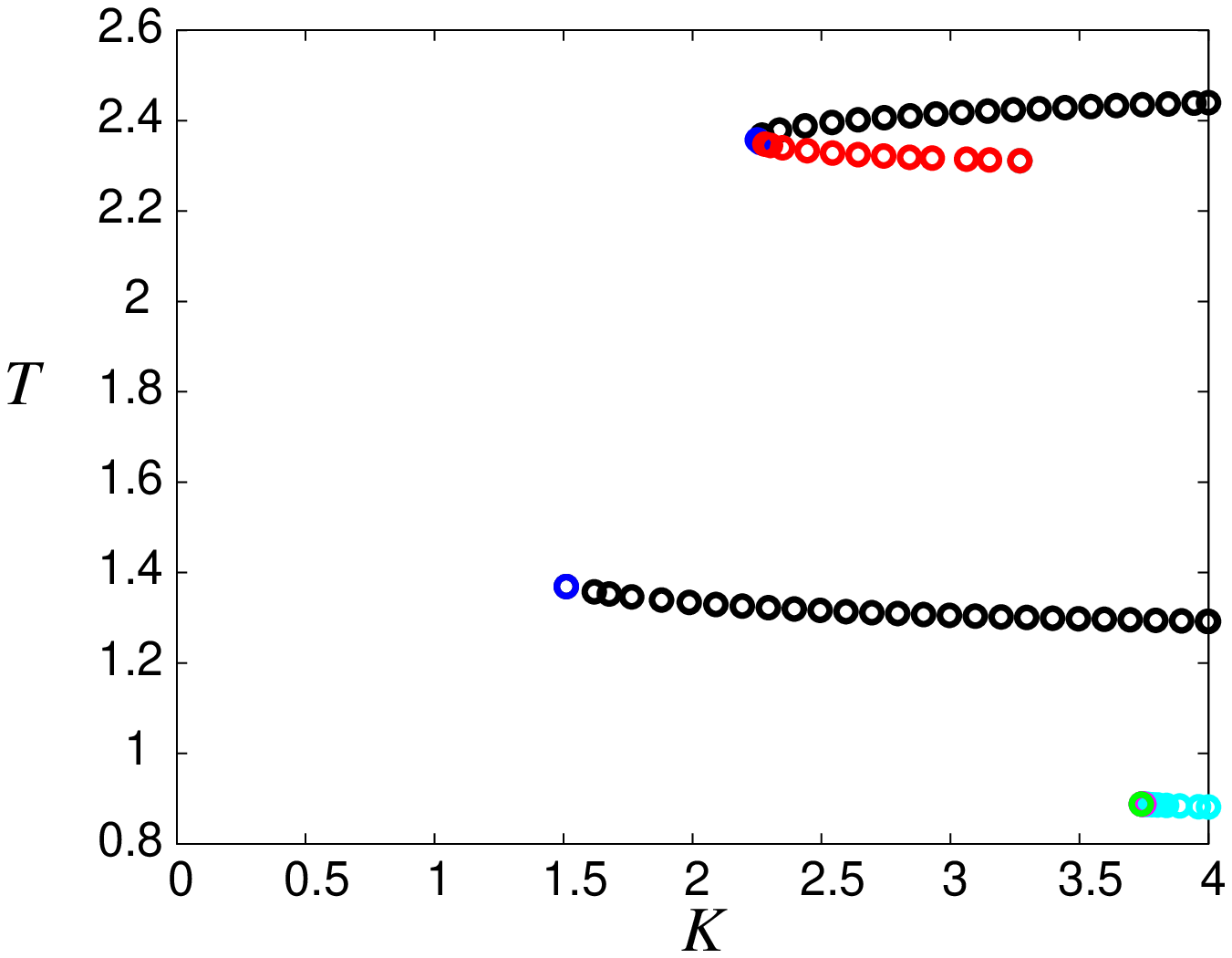}
\caption{(a)The bifurcation diagram of Eq.\eqref{2.8}. (b) The period of the bifurcating solutions of Eq.\eqref{2.8}. The parameters are $\omega_{0}=3$, $\delta=0.1$, $q_{0}=0.5$, $\gamma=0.5$ and $\tau=2.5$. Stable periodic orbits are labeled by black circles, unstable (1/2/3/4 Floquet exponents with positive real part) periodic orbits are labeled by blue/cyan/red/magenta circles,  respectively.}\label{fig2}\end{figure}

Similarly, if we fix $\tau=2.5$ and let $K$ vary, we obtain simulation results in Figure \ref{fig2}. We find that the system \eqref{2.8} exhibits incoherent state when $K$ is low and it exhibits partially coherent state when $K$ exceeds a precise threshold, which supports the obtained result in literature \cite{E. Montbrio1}. Moreover, if we further increase $k$ more than one branches of bifurcating solutions appear which may also be originated from a backward bifurcation (shown in  Figure \ref{fig2}(a) with red circles). It is well known that the system exhibits globally stable incoherent state when $K=0$, thus all branches of bifurcating solutions exists globally for $K\rightarrow\infty$. This means that  more than one branches of coherent states coexists (with similar order parameters shown in  Figure \ref{fig2}(a) with black circles) but have different periods (shown in  Figure \ref{fig2} (b)).

\begin{figure}\centering
a)
\includegraphics[width=0.46\textwidth]{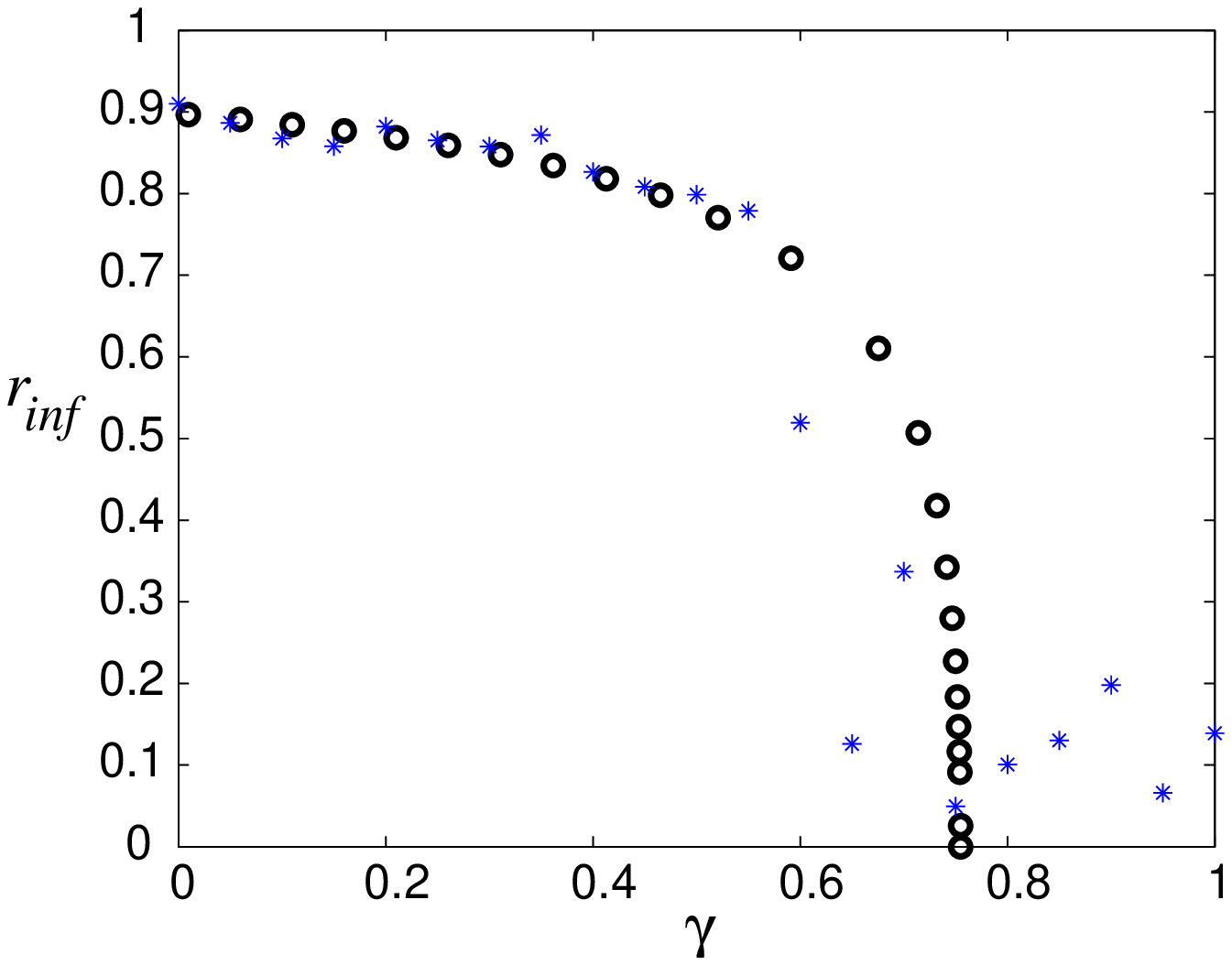}
b)
\includegraphics[width=0.46\textwidth]{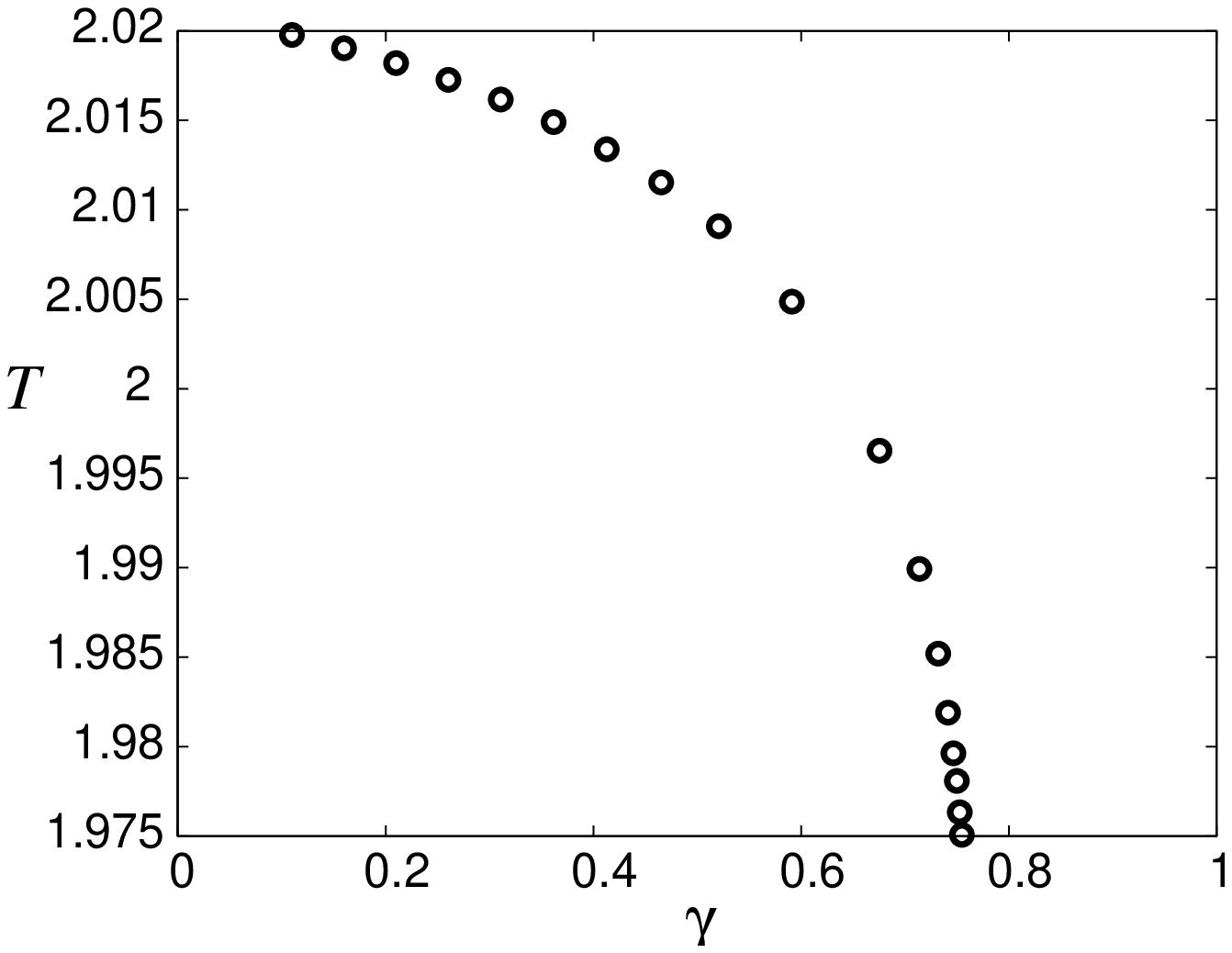}\\
c)
\includegraphics[width=0.46\textwidth]{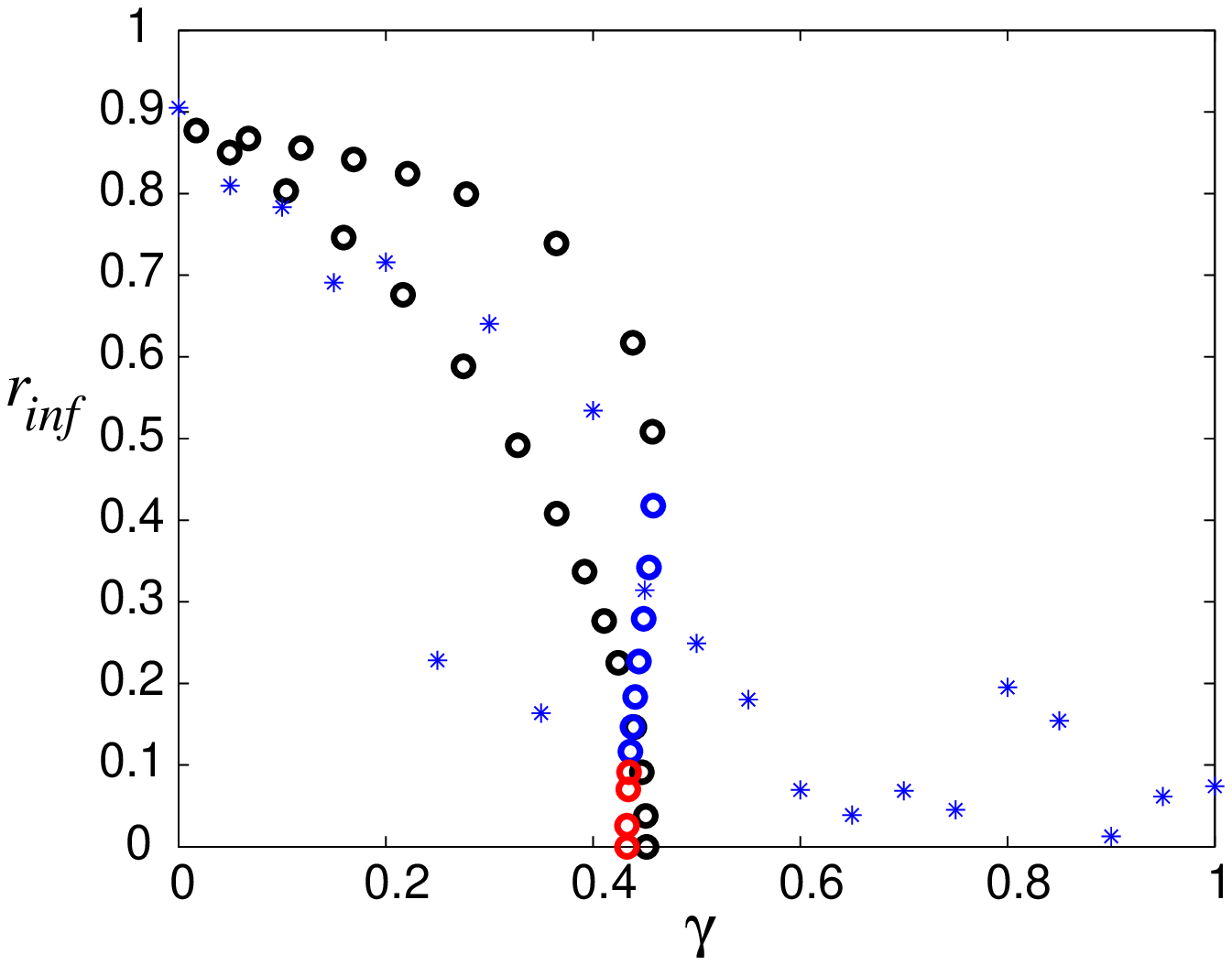}
d)
\includegraphics[width=0.46\textwidth]{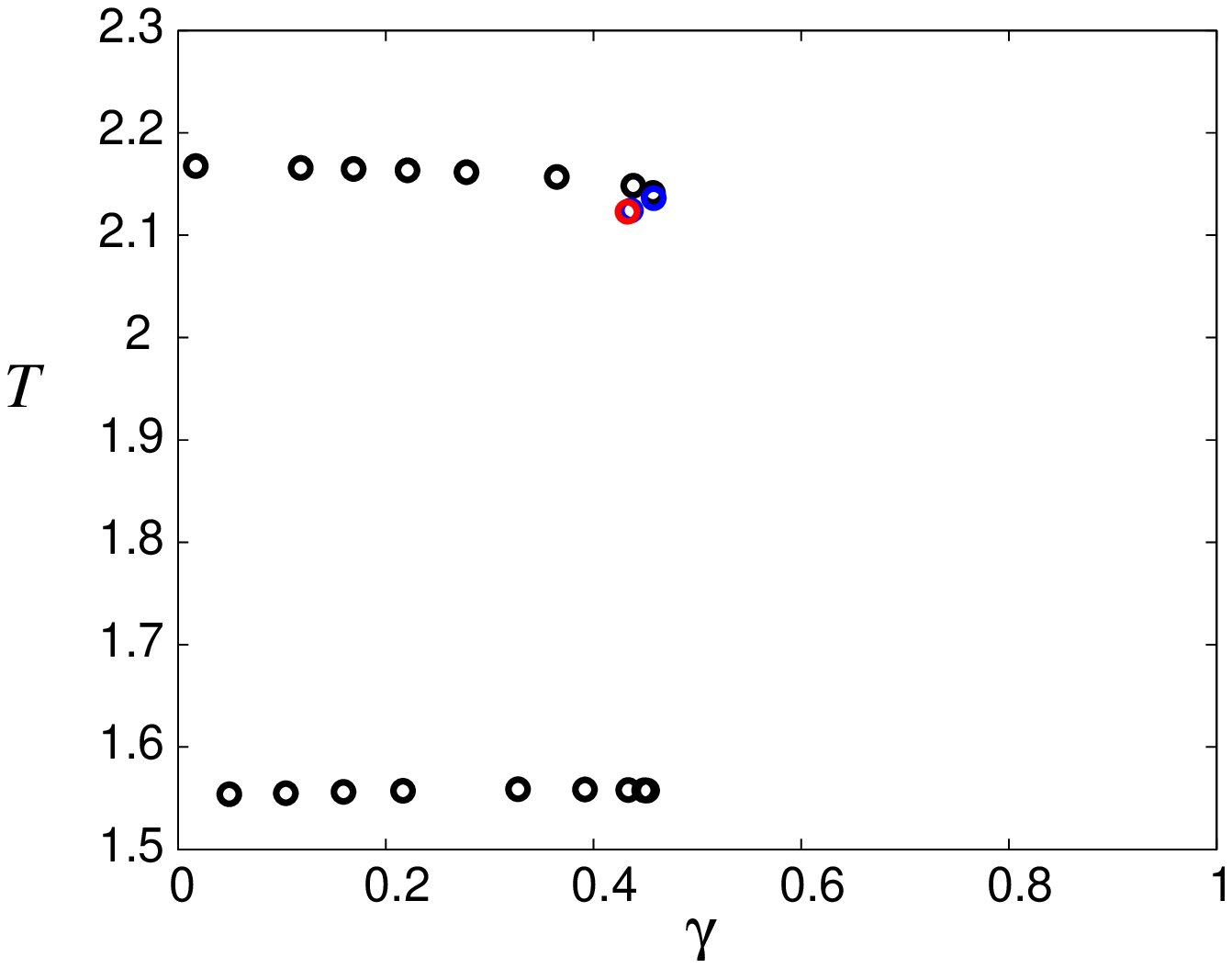}
\caption{The bifurcation diagram of Eq.\eqref{2.8} is shown in (a) $\tau=2$ and (c) $\tau=4.4$. The period of the bifurcating solutions is shown in (b) $\tau=2$ and (d) $\tau=4.4$.  $K=1$, $\omega_{0}=3$, $\delta=0.1$ and $q_{0}=0.5$. Stable periodic orbits are labeled by black circles, unstable (1/3 Floquet exponents with positive real part) periodic orbits are labeled by blue/red circles,  respectively.}\label{fig3}
\end{figure}

If we fix $\tau=2$ and $4.4$ respectively, and let $\gamma$ vary, we obtain simulation results in Figure \ref{fig3}.  We find that the system \eqref{2.8}
exhibits partially coherent state when $\gamma$ is low and it exhibits incoherent state when $\gamma$ exceeds a precise threshold. This indicates that large spread of shear will significantly eliminate the synchronization of the Kuramoto model, which is an extension of the results given in \cite{E. Montbrio1,E. Montbrio2} in the absence of delay. Larger  $\tau$ induces more branches of  coherent states as shown in these simulations, which may also be subcritical branches (red circles in Figure \ref{fig3} (c)).

\begin{figure}\centering
a)
\includegraphics[width=0.46\textwidth]{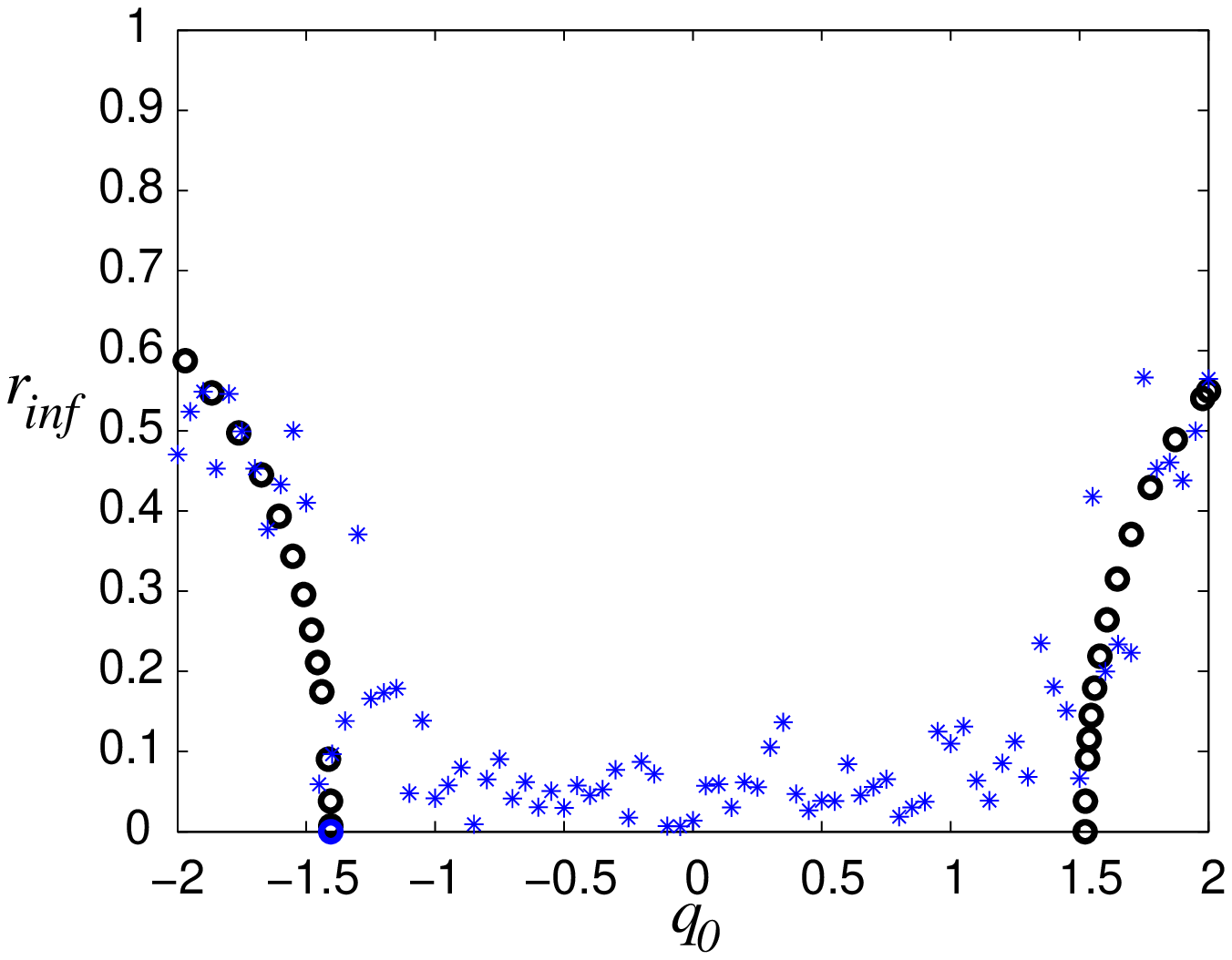}
b)
\includegraphics[width=0.46\textwidth]{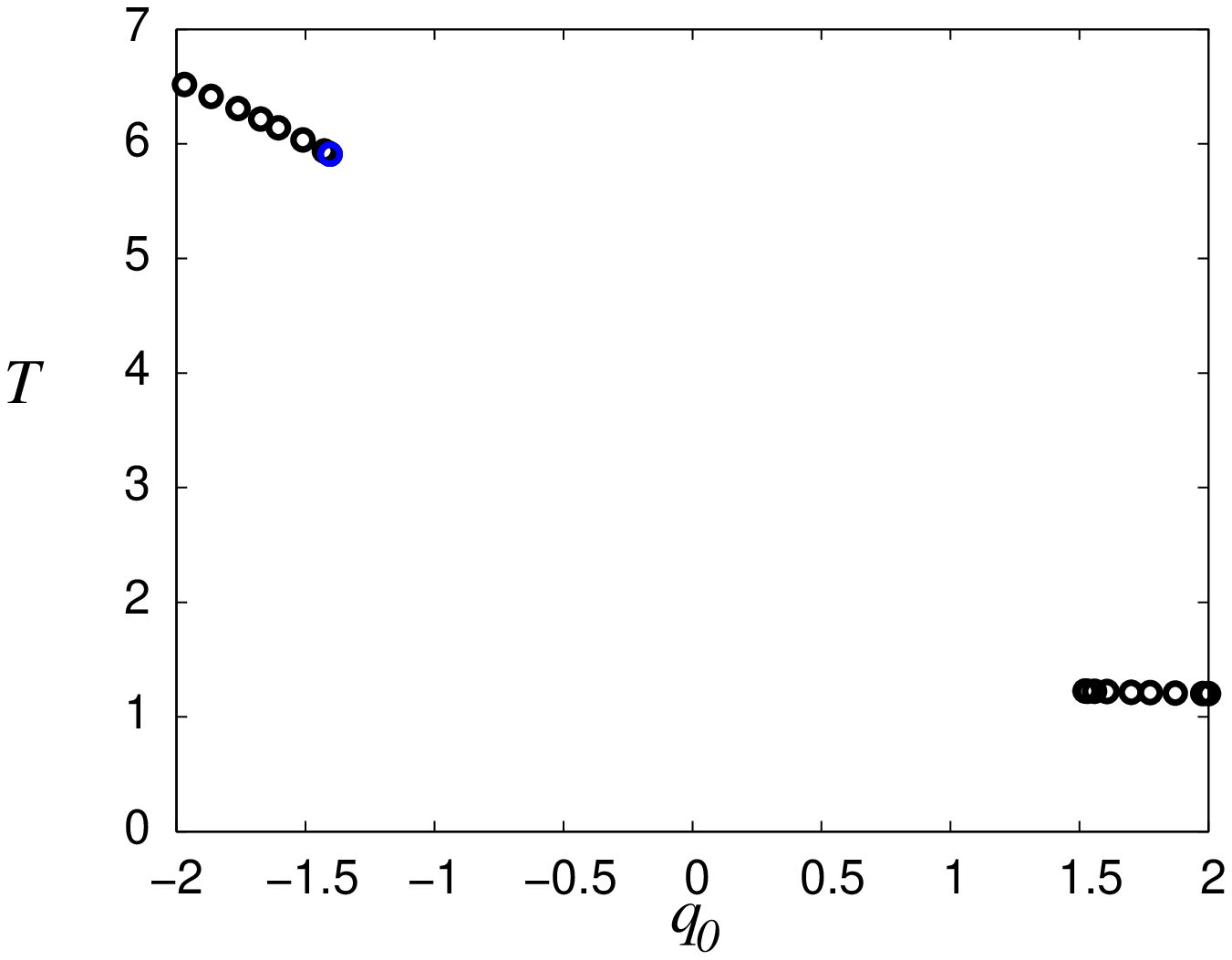}\\
c)
\includegraphics[width=0.46\textwidth]{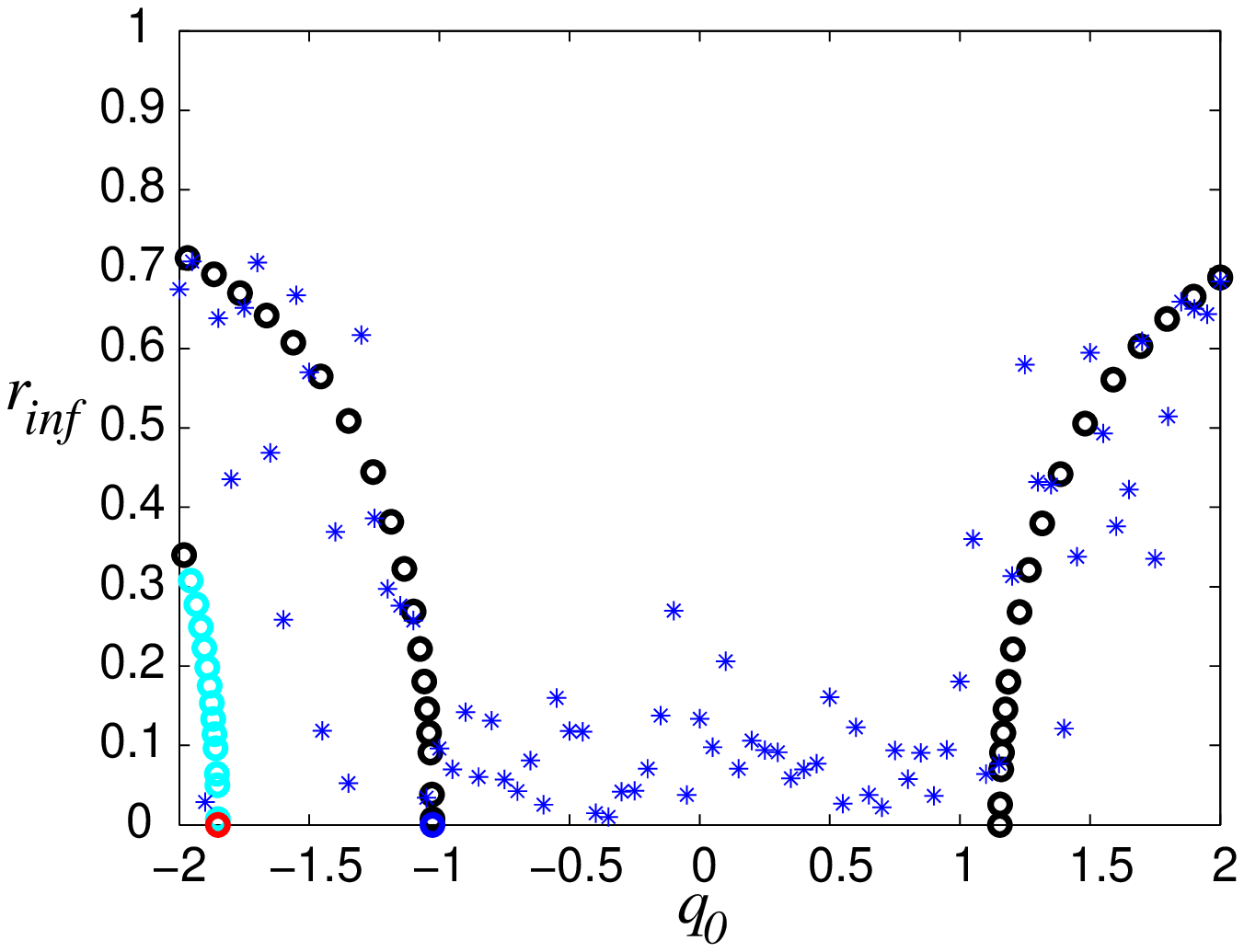}
d)
\includegraphics[width=0.46\textwidth]{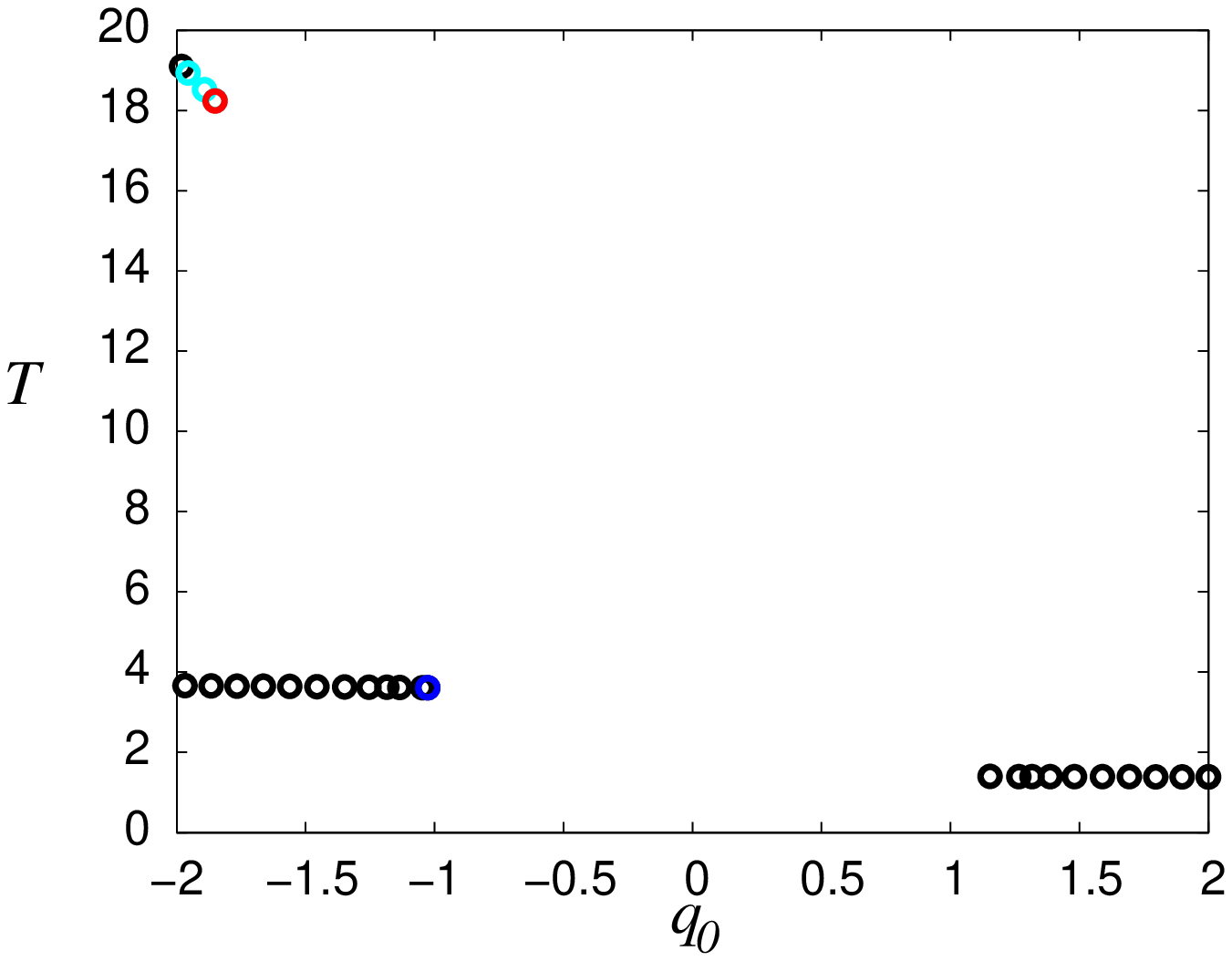}
\caption{The bifurcation diagram of Eq.\eqref{2.8} is shown in (a)$\tau=1$ and (c) $\tau=4$. The period of the bifurcating solutions of Eq.\eqref{2.8} is shown in b $\tau=1$ and (d) $\tau=4$.  $K=1$, $\omega_{0}=3$, $\delta=0.1$ and $\gamma=1$. Stable periodic orbits are labeled by black circles, unstable (2/3 Floquet exponents with positive real part) periodic orbits are labeled by cyan/red circles,  respectively.}\label{fig4}
\end{figure}

If we choose $K=1$, $\omega_{0}=3$, $\delta=0.1$, $\gamma=1$, $\tau=1$, and let $q_{0}$ vary, we obtain simulation results in Figure \ref{fig4}(a-b). We find that the system \eqref{2.8}
exhibits incoherent state when $|q_{0}|$ is low and it exhibits partially coherent state when $|q_{0}|$ exceeds a precise threshold. This indicates that large value of average shear induces partial synchronization in the Kuramoto model.   Using larger $\tau=4$, we give similar results in Figure \ref{fig4}(c-d), where we find this also brings more branches of coherent states. The two branches of coherent states are both stable when $\gamma$ is away from the bifurcation points, e.g., there exists one branch marked by red circles in Figure \ref{fig4}(c) which turns to be stable for small $\gamma$.

\subsection{Two-parameter bifurcations}
In this section, we will show the effect of shear and delay in two-parameter plane.

\begin{figure}\centering
a)
\includegraphics[width=1.8in]{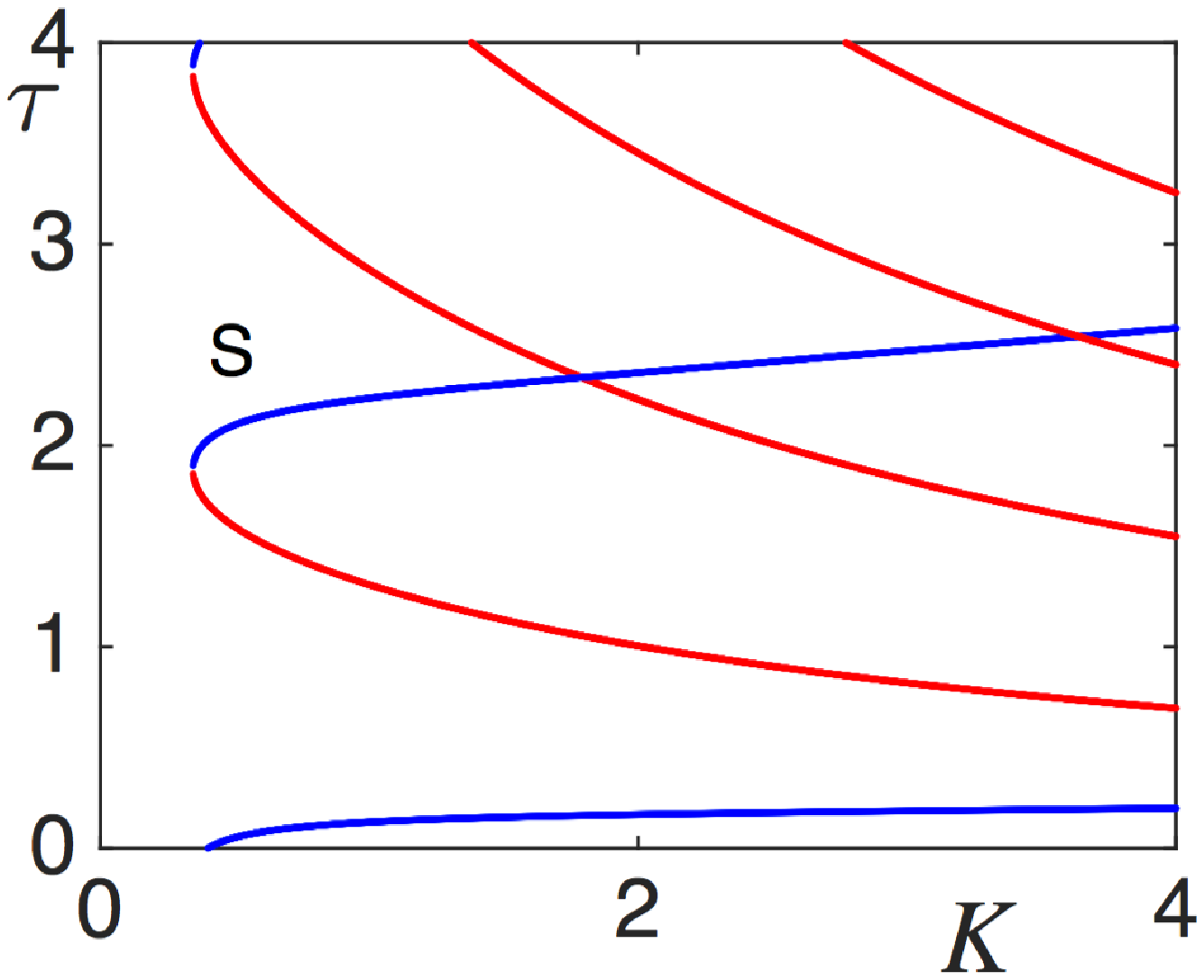}
b)
\includegraphics[width=1.8in]{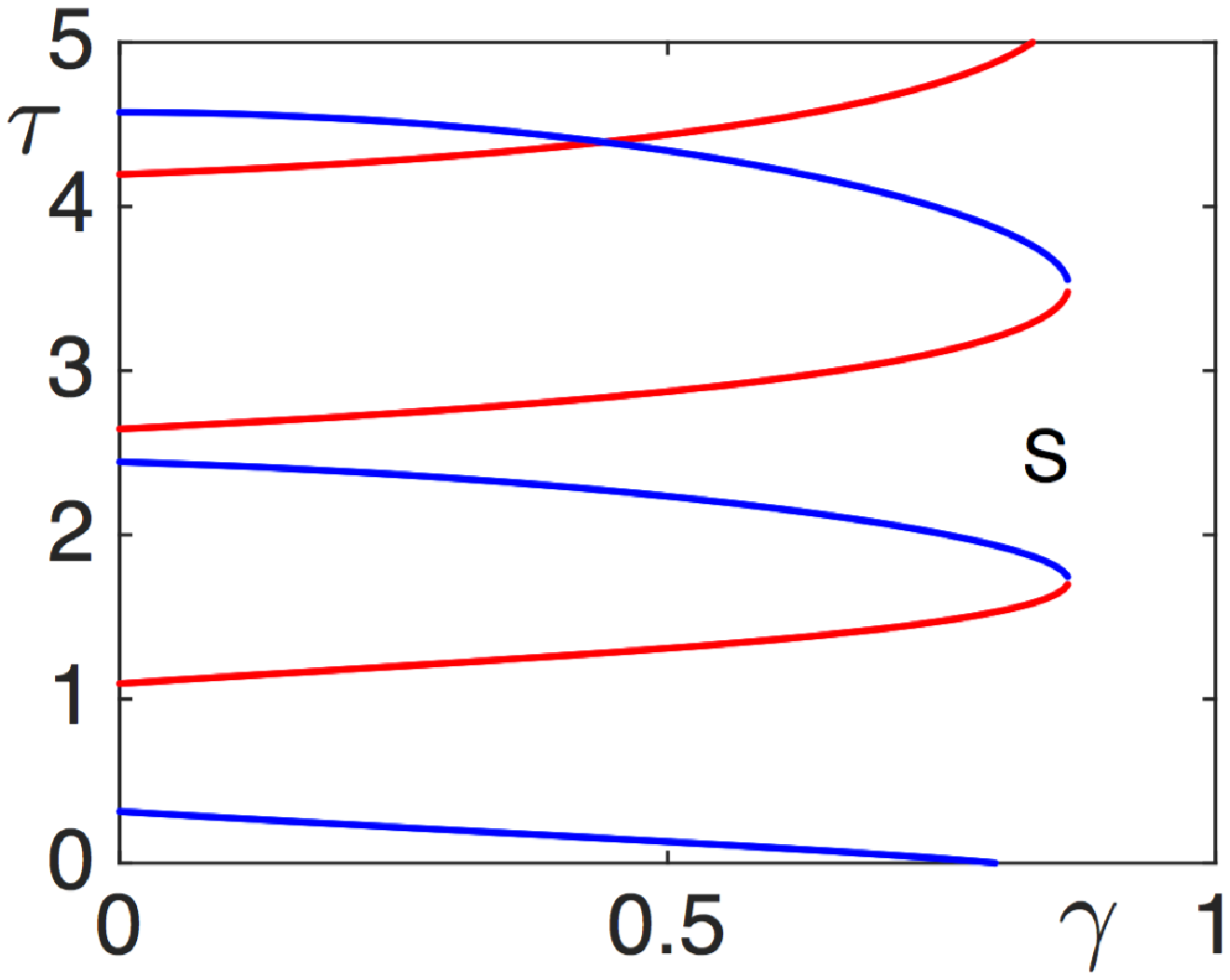}
c)
\includegraphics[width=1.8in]{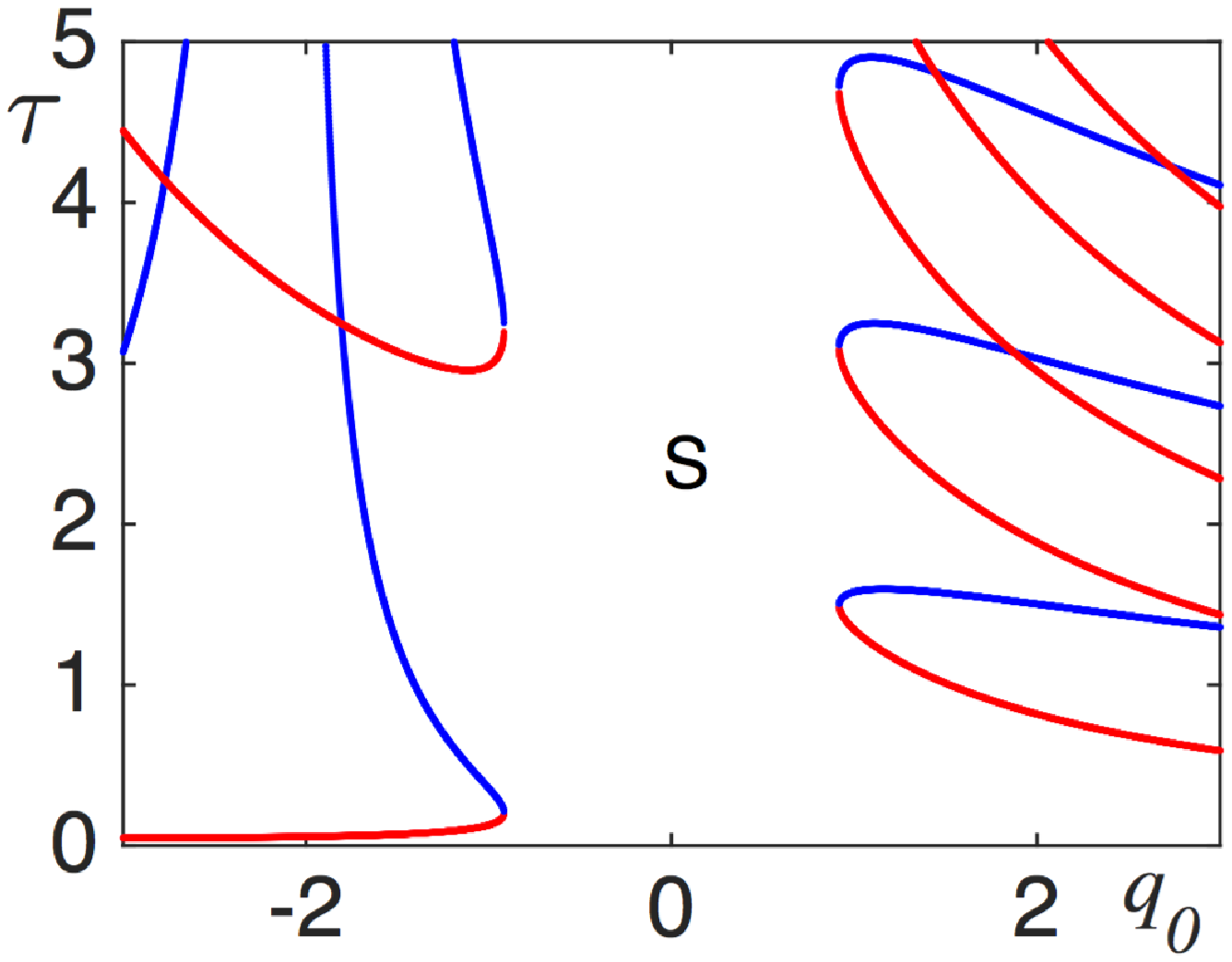}
\caption{Hopf bifurcations of Eq.\eqref{2.8} are shown. Regions marked by 'S' mean stable incoherent states exist. (a)$K-\tau$ plane for $\omega_0=3$, $\delta=0.1$, $q_0=0.5$ and $\gamma=0.5$. (b)$\gamma-\tau$ plane for $K=1$, $\omega_0=3$, $\delta=0.1$ and $q_0=0.5$. (c)$q_0-\tau$ plane for $K=1$, $\omega_0=3$, $\delta=0.1$ and $\gamma=1$. }\label{fig5}
\end{figure}

When $\omega_0=3$, $\delta=0.1$, $q_0=0.5$ and $\gamma=0.5$,   we draw the Hopf bifurcation values by curves shown in Figure \ref{fig5}(a). The red curves stands for $\tau_j^+$ and the blue curves stands for $\tau_j^-$. We can find that system \eqref{2.8} is incoherent when $K$ is less than the critical value and system is in partial synchronization when $K$ crosses the Hopf bifurcation values, which means the coupling strength enhances the synchronization. Also, the synchronization switches are also observed when varying time delay $\tau$.

When $K=1$, $\omega_0=3$, $\delta=0.1$ and $q_0=0.5$, we draw the Hopf bifurcation values by curves shown in Figure \ref{fig5}(b). The red curves stands for $\tau_j^+$ and the blue curves stands for $\tau_j^-$. We can find that system \eqref{2.8} is in partial synchronization when $\gamma$ is less than the critical value and system is incoherent when $\gamma$ crosses the Hopf bifurcation values, which means the large spread width of the shear can weaken the synchronization.
In this figure, we can clearly find that increasing $\tau$ leads to more branches of Hopf bifurcations which is also a theoretical explanation of Figure \ref{fig3}(a) and \ref{fig3}(c).

When $K=1$, $\omega_0=3$, $\delta=0.1$ and $\gamma=1$, we draw the Hopf bifurcation values by curves shown in Figure \ref{fig5} (c). The red curves stands for $\tau_j^+$ and the blue curves stands for $\tau_j^-$. We can find that system \eqref{2.8} is incoherent when $|q_0|$ is less than the critical value and system is in partial synchronization when $|q_0|$ crosses the Hopf bifurcation values, which means the absolute value of the mean of   shear can strengthen the synchronization. Again, as shown in this figure, we find  increasing $\tau$ leads to more branches of Hopf bifurcations.

\subsection{Three-parameter bifurcations}

\begin{figure}\centering
\centering
a)\includegraphics[width=0.4\textwidth]{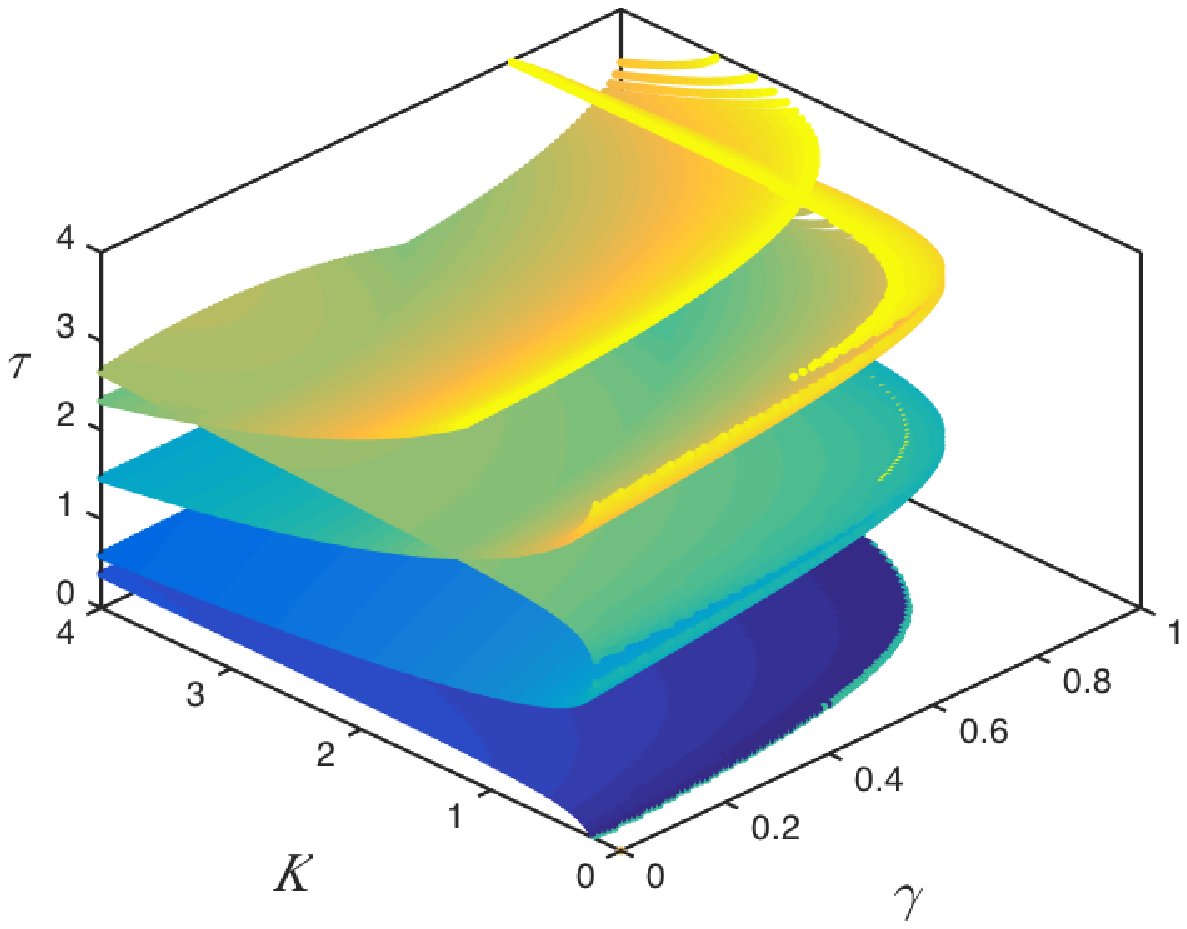}b)\includegraphics[width=0.4\textwidth]{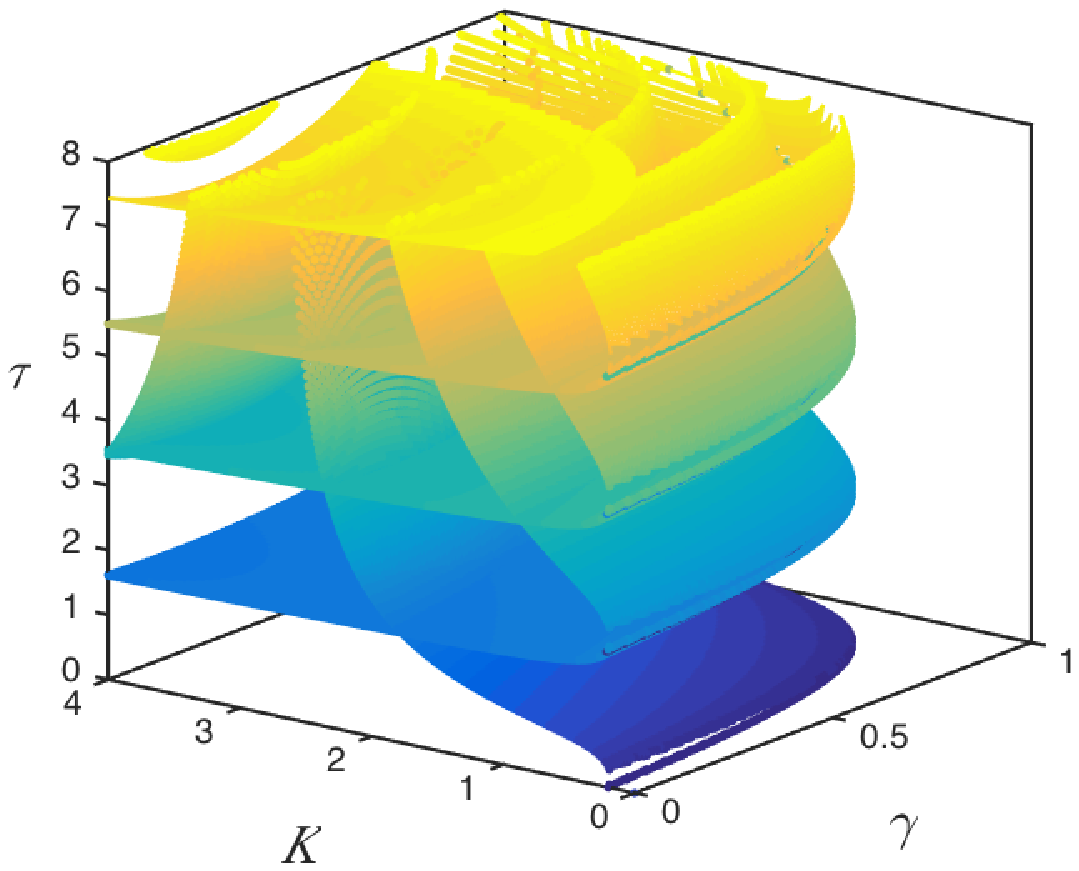}
\caption{Hopf bifurcation surfaces of Eq.\eqref{2.8} are shown when  $\omega_0=3$, $\delta=0.1$ and (a) $q_0=0.5$ (b) $q_0=-0.5$. When parameters crosses the surfaces along the direction that $\gamma$ decreases, the incoherent state looses its stability and coherent states appear.}\label{fig66}
\end{figure}

When $\omega_0=3$, $\delta=0.1$ and $q_0=0.5$, we draw the Hopf bifurcation values by surfaces shown in Figure \ref{fig66}. In this figure, we combine the previous one-parameter or two-parameter bifurcation results.
 As discussed in the previous section, we find that the incoherent state loses its stability and coherent states appear, when parameters crosses the surfaces along the direction that $\gamma$ decreases. We can find that system \eqref{2.8} is in partial synchronization when $\gamma$ is less than the critical value and system is incoherent when $\gamma$ crosses the Hopf bifurcation values, which means the spread width of the shear weaken the synchronization. Moreover, in both figures, one can find synchronization windows when time delay $\tau$ increases. The number of the resonant structures, i.e., the number of synchronization windows, becomes fewer when the coupling strength $K$ or spread of shear $\gamma$ increases. When $\gamma$, the spread of shear, is small, the effect of using positive of negative $q_0$ the mean value of shear is not obvious. When $\gamma$ is large, the effect is obvious: in case of negative $q_0$, there is only one stable region for time delay $\tau$.

\begin{figure}\centering
\centering
a)\includegraphics[width=0.4\textwidth]{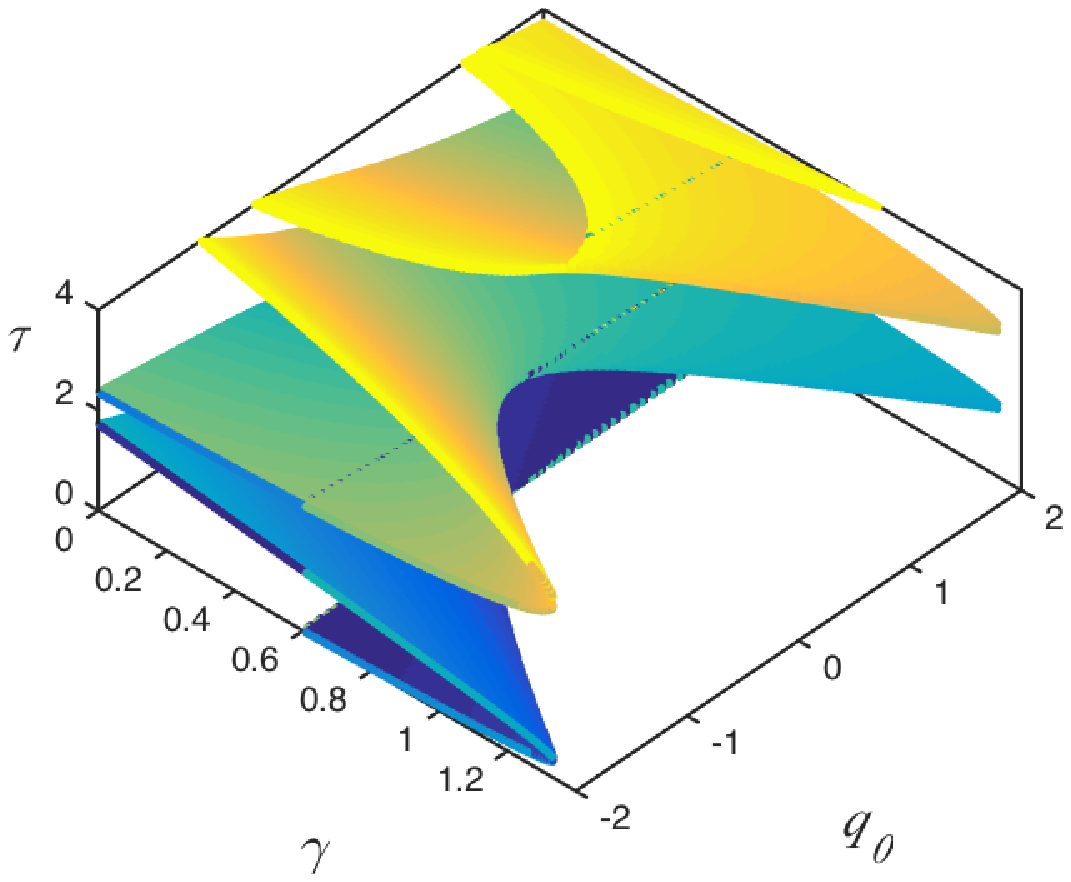}b)\includegraphics[width=0.4\textwidth]{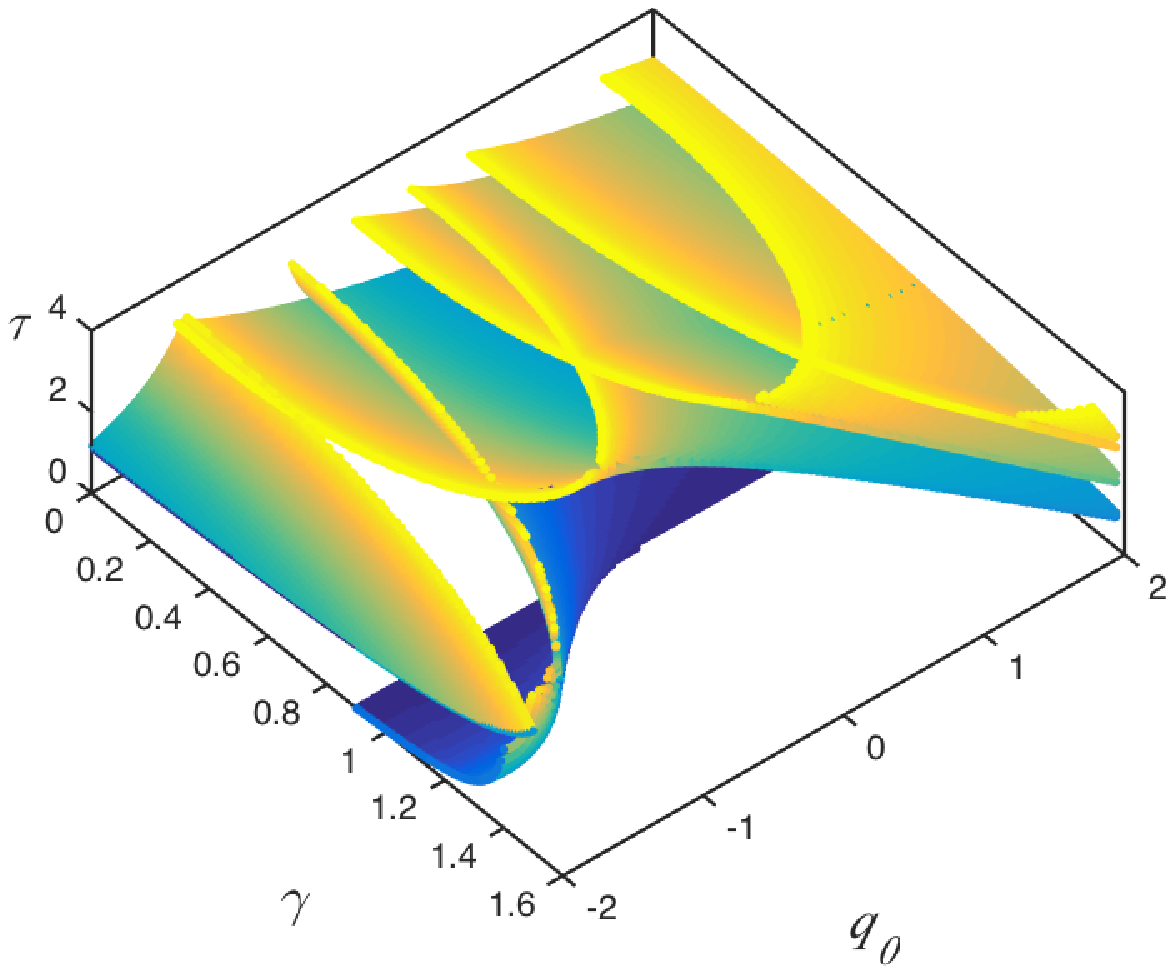}
\caption{Hopf bifurcation surfaces of Eq.\eqref{2.8} are shown when  $\omega_0=3$, $\delta=0.1$ and (a) $K=0.5$ (b) $K=2$. When parameters crosses the surfaces along the direction that $\gamma$ decreases, the incoherent state looses its stability and coherent states appear.}\label{fig67}
\end{figure}

 For fixed $K$, we can similarly discuss the effect of delay and shear in Figure \ref{fig67}.  As shown in figure (a), clearly  we have  the system \eqref{2.8}
exhibits incoherent state when $\gamma$ is large and  $|q_{0}|$ is small. The Kuramoto model exhibits partially coherent states when $|q_{0}|$ exceeds a precise threshold for large $\gamma$. When $\gamma$ is small, that means small inhomogenerity of  oscillators, the Kuramoto model exhibits stable incoherent state when $\tau$ in some particularly interval, as shown in the figure \ref{fig67} (a) the surfaces are separate from each other.  For a large coupling strength $K=2$ in figure \ref{fig67} (b), the situation is different, we find that the effect of  both the delay and the shear is weak. As shown in figure (b), we find that the Kuramoto model exhibits stable coherent states for any $\tau$ when $\gamma$ is small. From these two simulations, we can also conclude that for large coupling strength $K$, the parameter region standing for incoherent state is also large.

\section{Conclusion}
In this paper, we analyzed the influence of time delay and distributed shear on the dynamics of Kuramoto model, from the point of view of the bifurcation analysis on the Ott-Antonsen's manifold. Mainly, three bifurcation parameters are investigated, time delay $\tau$, the mean value of shear $q_0$ and the spread of shear $\gamma$. The sufficient conditions to ensure asymptotical stability of the incoherence and the existence of Hopf bifurcations, which determines the synchronization transition, are both obtained by analyzing the distribution of the eigenvalues. In addition, we established the explicit formula by deriving   normal forms of Hopf bifurcations to determine the direction and stability of periodic solutions bifurcating from the incoherence.

We found that increasing time delays would lead the Kuramoto model to synchronization switching.  On one hand, time delay may induce synchronization thus lead the system to order. On the other, time delay can eliminate synchronization  thus lead the system to disorder (see Theorem \ref{lem3.3}). The effect of mean value and variance of the shears on the dynamics are also obtained. As expected, decreasing the variance will lead the system to coherent states. Also, increasing the mean value will lead the system to coherent states.  Because of the existence of time delay, there may exist more than one branches of bifurcated coherent states. Some local and global bifurcation diagram are given in multiple-parameter spaces, which indicate large  time delay and suitable shears could induces  several stable coherent states in Kuramoto model simultaneously.

\section*{Acknowledgements}
The authors wish to express their special gratitude to the editors and the reviewers for the helpful comments given for this paper. This research  is supported by NNSFC (11701120).

\section*{Appendix}
In this Appendix we will calculate the coefficients which determines  bifurcation properties listed in Theorem \ref{the4.1}.

Setting $r^*(t)=x(t)+{\rm i}y(t)$, in Eq.\eqref{2.8}, and separating the real and imaginary parts, we get the equivalent system
\begin{equation}
\label{4.1}
\begin{aligned}
\dot{x}(t)=&-(\delta+K\gamma)x(t)+({\omega}_{0}+K{q}_{0})y(t)+\frac{K}{2}(\gamma+1)x(t-\tau)-\frac{K{q}_{0}}{2}y(t-\tau)\\
        &+\frac{K}{2}(\gamma-1){x}^{2}(t)x(t-\tau)+\frac{K}{2}(1-\gamma){y}^{2}(t)x(t-\tau)+K(\gamma-1)x(t)y(t)y(t-\tau)\\
        &+\frac{K{q}_{0}}{2}{x}^{2}(t)y(t-\tau)-\frac{K{q}_{0}}{2}{y}^{2}(t)y(t-\tau)-K{q}_{0}x(t)y(t)x(t-\tau)\\
\dot{y}(t)=&-({\omega}_{0}+K{q}_{0})x(t)-(\delta+K\gamma)y(t)+\frac{K{q}_{0}}{2}x(t-\tau)+\frac{K}{2}(\gamma+1)y(t-\tau)\\
        &+\frac{K}{2}(1-\gamma){x}^{2}(t)y(t-\tau)
        +\frac{K}{2}(\gamma-1){y}^{2}(t)y(t-\tau)+K(\gamma-1)x(t)y(t)x(t-\tau)\\
        &+K{q}_{0}x(t)y(t)y(t-\tau)+\frac{K{q}_{0}}{2}{x}^2(t)x(t-\tau)-\frac{K{q}_{0}}{2}{y}^{2}(t)x(t-\tau)
\end{aligned}
\end{equation}
We use $a=-(\delta+K\gamma)$, $b=\frac{K}{2}(1+\gamma)$, $c={\omega}_{0}+K{q}_{0}$ and $d=\frac{K{q}_{0}}{2}$ to avoid very long expressions.
The characteristic equation associated with the linearization of \eqref{4.1}, around the incoherence $(x,y)=(0,0)$, is,
\begin{equation}
\label{4.2}
(\lambda-a-b{\rm e}^{-\lambda\tau})^2+(c-d{\rm e}^{-\lambda\tau})^2=0
\end{equation}
Now we can rescale the time by $t\mapsto(t/\tau)$ to normalize the delay so that system \eqref{4.1} can be written in the following form
\begin{equation}
\label{4.3}
\begin{aligned}
\dot{x}(t)=&-\tau(\delta+K\gamma)x(t)+\tau({\omega}_{0}+K{q}_{0})y(t)+\frac{K\tau}{2}(\gamma+1)x(t-1)-\frac{K{q}_{0}\tau}{2}y(t-1)\\&+\frac{K\tau}{2}(\gamma-1){x}^{2}(t)x(t-1)
           +\frac{K\tau}{2}(1-\gamma){y}^{2}(t)x(t-1)+K\tau(\gamma-1)x(t)y(t)y(t-1)\\&+\frac{K{q}_{0}\tau}{2}{x}^{2}(t)y(t-1)-\frac{K{q}_{0}\tau}{2}{y}^{2}(t)y(t-1)-K{q}_{0}\tau x(t)y(t)x(t-1)\\
\dot{y}(t)=&-({\omega}_{0}+K{q}_{0})\tau x(t)-(\delta+K\gamma)\tau y(t)+\frac{K{q}_{0}\tau}{2}x(t-1)+\frac{K\tau}{2}(\gamma+1)y(t-1)\\&+\frac{K\tau}{2}(1-\gamma){x}^{2}(t)y(t-1)+\frac{K\tau}{2}(\gamma-1){y}^{2}(t)y(t-1)\\&+K\tau(\gamma-1)x(t)y(t)x(t-1)+K{q}_{0}\tau x(t)y(t)y(t-1)+\frac{K{q}_{0}\tau}{2}{x}^2(t)x(t-1)-\frac{K{q}_{0}\tau}{2}{y}^{2}(t)x(t-1)
\end{aligned}
\end{equation}
The characteristic equation of the linearization of \eqref{4.3}, around the incoherence $(x,y)=(0,0)$, is,
\begin{equation}
\label{4.4}
(z-\tau a-\tau b {\rm e}^{-z})^2+(\tau c-\tau d{\rm e}^{-z})^2=0
\end{equation}
Comparing Eq.\eqref{4.4} with Eq.\eqref{4.2}, one can find that $z=\lambda\tau$. So by the result in Section 3, we have that Eq.\eqref{4.4} has a pair of pure imaginary roots $\pm {\rm i}\tau_j^\pm \beta_\pm$. Let $z(\tau)$ be the root of Eq.\eqref{4.4} satisfying $\mathrm{Re}z(\tau_j^\pm)=0$ and $\mathrm{Im}z(\tau_j^\pm)=\tau_j^\pm\beta_\pm$.

Clearly, the phase space is $C=C([-1,0],R^2)$. For convenience, denote $\tau=\overline{\tau}+\mu$ for $\overline{\tau}\in{\tau_j^\pm}$ and $\mu\in R$. Then $\mu=0$ is the Hopf bifurcation value of system \eqref{4.3}. Let $z(\tau)$ be the root of Eq.\eqref{4.4} when $\tau=\overline{\tau}$, where either $\beta_0=\beta_+$ or $\beta_0=\beta_-$.

   For $\phi=(\phi_1,\phi_2)\in C$, let $L_\mu\varphi=(\overline{\tau}+\mu)A\phi(0)+(\overline{\tau}+\mu)B\phi(-1)$, where
\begin{equation*}
A=
\begin{pmatrix}
a  & c \\
-c & a
\end{pmatrix}
,B=
\begin{pmatrix}
b  & -d \\
d  & b
\end{pmatrix}
\end{equation*}
and
\begin{equation}
\label{4.6}
\begin{split}
f(\mu,\phi)=&(\overline{\tau}+\mu)\big[
\begin{pmatrix}
\frac{K}{2}(\gamma-1)x^2(t)x(t-1)+\frac{K}{2}(1-\gamma)y^2(t)x(t-1) \\
\frac{K}{2}(1-\gamma)x^2(t)y(t-1)+\frac{K}{2}(\gamma-1)y^2(t)y(t-1)
\end{pmatrix}
\\&+
\begin{pmatrix}
K(\gamma-1)x(t)y(t)y(t-1)+\frac{Kq_{0}}{2}x^2(t)y(t-1) \\
K(\gamma-1)x(t)y(t)x(t-1)+Kq_0x(t)y(t)y(t-1)
\end{pmatrix}\\&+
\begin{pmatrix}
-\frac{Kq_{0}}{2}y^2(t)y(t-1)-Kq_{0}x(t)y(t)x(t-1) \\
\frac{Kq_0}{2}x^2(t)x(t-1)-\frac{Kq_0}{2}y^2(t)x(t-1)
\end{pmatrix}
\big]
\end{split}
\end{equation}
By the Riese representation theorem, there exists matrix whose components are bounded variation function $\eta(\theta,\mu)$ in $\theta\in[-1,0]$ such that
\begin{equation*}
{L}_{\mu}\phi={\int}_{-1}^{0}{\rm d}\eta(\theta,\mu)\phi(\theta),\quad\forall \phi\in C
\end{equation*}
where bounded variation function $\eta(\theta,\mu)$ can be chosen as
\begin{equation*}
\eta(\theta,\mu)=(\overline{\tau}+\mu)A\delta(\theta)-(\overline{\tau}+\mu)B\delta(\theta+1)
\end{equation*}
where
\begin{equation*}
\delta(\theta)=
\begin{cases}
1,\theta=0 \\
0,\theta\neq0
\end{cases}
\end{equation*}
For $\phi\in C^1([-1,0],R^2)$, define
\begin{equation}
\label{4.7}
A(\mu)\phi=
\begin{cases}
\frac{{\rm d}\phi(\theta)}{{\rm d}\theta},& \theta\in[-1,0),\\
\int_{-1}^{0}{\rm d}\eta(\eta,s)\phi(s),  & \theta=0,
\end{cases}
\end{equation}
and
\begin{equation}
\label{4.8}
R(\mu)\phi=
\begin{cases}
0,          & \theta\in[-1,0),\\
f(\mu,\phi),& \theta=0,
\end{cases}
\end{equation}
The system \eqref{4.3} is equivalent to
\begin{equation}
\label{4.9}
\dot{{u}_{t}}=A(\mu){u}_{t}+R(\mu){u}_{t}
\end{equation}

For $\psi\in C^1([0,1],(R^2)^*)$, define
\begin{equation}
\label{4.10}
{A}^{*}\psi(s)=
\begin{cases}
-\frac{{\rm d}\psi(s)}{{\rm d}s},             & s\in(0,1],\\
{\int}_{-1}^{0}{\rm d}\eta(t,0)\psi(-t),      & s=0,
\end{cases}
\end{equation}
and a bilinear inner product
\begin{equation}
\label{4.11}
\langle\psi(s),\phi(\theta)\rangle=\overline{\psi}(0)\phi(0)-\int_{-1}^{0}\int_{\xi=0}^{\theta}\overline{\psi}(\xi-\theta){\rm d}\eta(\theta)\phi(\xi){\rm d}\xi,
\end{equation}
where $\eta(\theta)=\eta(\theta,0)$. Then $A(0)$ and $A^*$ are adjoint operators. In addition, from Section 3 we know that $\pm {\rm i}\overline{\tau}\beta_0$ are eigenvalues of $A(0)$. Thus, they are also eigenvalues of $A^*$. Let $q(\theta)=(m,1)^T{\rm e}^{{\rm i}\overline{\tau}\beta_0\theta}$ is the eigenvector of $A(0)$ corresponding to ${\rm i}\overline{\tau}\beta_0$ and $q^*(s)=\frac{1}{\overline{D}}(n,1){\rm e}^{{\rm i}\overline{\tau}\beta_0s}$ is the eigenvector of $A^*$ corresponding to $-{\rm i}\overline{\tau}\beta_0$. Then it is not difficult to show that
\begin{equation*}
m=n=i
\end{equation*}
Thus, using $\langle q^*(s),q(\theta)\rangle=1$, we have
\begin{equation*}
\begin{split}
\langle{q}^{*}(s),q(\theta)\rangle&=\overline{q}^{*}(0)q(0)-\int_{-1}^{0}\int_{0}^{\theta}\overline{q}^{*}(\xi-\theta){\rm d}\eta(\theta)q(\xi){\rm d}\xi\\
                                  &=\frac{1}{D}\Big[(1+\overline{n}m)-(\overline{n},1)L\begin{pmatrix}\theta{\rm e}^{{\rm i}\overline{\tau}{\beta}_{0}\theta}\\\theta{\rm e}^{{\rm i}\overline{\tau}{\beta}_{0}\theta }m\end{pmatrix}\Big]
\end{split}
\end{equation*}
So that, we can get
\begin{equation}
\label{4.12}
\begin{split}
D=2+K\tau{\rm e}^{-{\rm i}\overline{\tau}\beta_{0}}(1+\gamma+q_{0}{\rm i})
\end{split}
\end{equation}

Using the same notations at in Hassard et al \cite{B.D. Hassard}, we compute the coordinates to describe the center manifold $C_0$. Let $u_t$ be the solution of Eq.\eqref{4.3} when $\mu=0$.\\
Define
\begin{equation}
\label{4.13}
z(t)=<q^*,u_t>,\quad W(t,\theta)=u_t(\theta)-2\mathrm{Re}{z(t)q(\theta)}.
\end{equation}
On the center manifold $C_0$, we have
\begin{equation*}
W(t,\theta)=W(z(t),\overline{z}(t),\theta),
\end{equation*}
where
\begin{equation}
\label{4.14}
W(z,\overline{z},\theta)=W_{20}(\theta)\frac{z^2}{2}+W_{11}(\theta)z\overline{z}+W_{02}(\theta)\frac{\overline{z}^2}{2}+\cdots
\end{equation}
$z$ and $\overline{z}$ are local coordinates for center manifold $C_0$ in the direction of $q^*$ and ${\overline{q}}^*$. Note that $W$ is real if $u_t$ is real. We consider only real solutions. For solution $u_t\in C_0$ of Eq.\eqref{4.3}, since $\mu=0$, we have
\begin{equation}
\label{4.15}
\begin{split}
\dot{z}(t)&={\rm i}\beta\overline{\tau}z+\overline{q}(\theta) f(0,\beta(z,\overline{z},\theta)+2{\mathrm Re}{zq(\theta)}) \\
          &={\rm i}\beta\overline{\tau}z+\overline{q}^*(0) f_0,
\end{split}
\end{equation}
that is
\begin{equation}
\label{4.16}
\dot{z}(t)={\rm i}\beta\overline{\tau}z(t)+g(z,\overline{z}),
\end{equation}
where
\begin{equation}
\label{4.17}
g(z,\overline{z})=g_{20}\frac{z^2}{2}+g_{11}z\overline{z}+g_{02}\frac{\overline{z}^2}{2}+g_{21}\frac{z^2\overline{z}}{2}+\cdots.
\end{equation}
Then it follows from Eq.\eqref{4.13} that
\begin{equation}
\label{4.18}
\begin{split}
u_t(\theta)=&W(t,\theta)+2\mathrm{Re}{z(t)q(\theta)} \\
           =&W_{20}(\theta)\frac{z^2}{2}+W_{11}(\theta)z\overline{z}+W_{02}\frac{\overline{z}^2}{2}+(i,1)^T{\rm e}^{{\rm i}\beta_0\overline{\tau}\theta}z+(-i,1)^T{\rm e}^{-{\rm i}\beta_0\overline{\tau}\theta}\overline{z}+\cdots
\end{split}
\end{equation}
Combining with Eq.\eqref{4.6}, Eq.\eqref{4.17} and Eq.\eqref{4.18} and comparing the coefficients, we obtain
\begin{equation*}
\begin{split}
g_{20}=&g_{11}=g_{02}=0,{}\\
g_{21}=&\frac{8K\tau{\rm e}^{{\rm i}\beta_{0}\tau}}{D}(\gamma-1+q_{0}{\rm i})=\frac{8K\tau{\rm e}^{{\rm i}\beta_{0}\tau}(\gamma-1+q_{0}{\rm i})}{2+K\tau{\rm e}^{-{\rm i}\overline{\tau}\beta_{0}}(1+\gamma+q_{0}{\rm i})}
\end{split}
\end{equation*}
Thus, we can compute the following values by the method given in \cite{B.D. Hassard}.
\begin{equation*}
\begin{split}
&c_1(0)=\frac{\rm i}{2\beta_0}[g_{11}g_{20}-2|g_{11}|^2-\frac{|g_{02}|^2}{3}]+\frac{g_{21}}{2}\\
&\mu_2=-\frac{\mathrm{Re}c_1(0)}{\alpha^{'}(\overline{\tau})}\\
&\beta_2=2\mathrm{Re}c_1(0)\\
&\tau_2=-\frac{\mathrm{Im}(c_1(0))+\mu_2\beta^{'}(\overline{\tau})}{\beta_0}
\end{split}
\end{equation*}

\bibliographystyle{model1-num-names}

\end{document}